\documentclass[hyper,11pt,paper]{article}
\usepackage[centertags]{amsmath}
\usepackage{amssymb}
\usepackage[english]{babel}
\usepackage{jheppub}
\usepackage{physics}

\newcommand{\cA}{\mathcal{A}}
\newcommand{\cB}{\mathcal{B}}
\newcommand{\cD}{\mathcal{D}}
\newcommand{\cE}{\mathcal{E}}
\newcommand{\cH}{\mathcal{H}}
\newcommand{\cI}{\mathcal{I}}
\newcommand{\cL}{\mathcal{L}}

\newcommand{\cO}{\mathcal{O}}

\newcommand{\cR}{\mathcal{R}}
\newcommand{\rd}{\mathrm{d}}

\newcommand{\secref}[1]{section~\ref{#1}}

\title{Comments on entanglement entropy in the $c=1$ string theory}

\author[a]{Ben Craps,}
\author[a]{Marius Gerbershagen,}
\author[a]{Maxim Pavlov,}
\author[b]{Alejandro Vilar López}
\affiliation[a]{Theoretische Natuurkunde, Vrije Universiteit Brussel (VUB) and The International Solvay Institutes, Pleinlaan 2, B-1050 Brussels, Belgium}
\affiliation[b]{Department of Physics and Astronomy, University of British Columbia,
6224 Agricultural Road, Vancouver, B.C. V6T 1Z1, Canada}
\emailAdd{Ben.Craps@vub.be}
\emailAdd{marius.gerbershagen@vub.be}
\emailAdd{Maxim.Dmitrievich.Pavlov@vub.be}
\emailAdd{alejandro.vilarlopez@ubc.ca}

\abstract{We study entanglement entropy in the low-energy effective field theory of two-dimensional string theory as well as in the singlet sector of the dual $c=1$ matrix quantum mechanics. From the target space perspective, we argue that a generic bulk subregion is expected to have an associated generalized entanglement entropy combining a dilaton-dependent gravitational term and a matter contribution coming from the tachyon, extending the construction from black hole horizons to arbitrary subregions. Motivated by the absence of the gravitational area-like term in previous analyses of entanglement entropy in the $c=1$ model, we 
compute the effect on the entanglement entropy of the nonlocal transformation induced by the leg-pole factor that relates the target space tachyon and the matrix model collective excitations, finding that it cannot account for the area-like term. We offer preliminary comments on the possible origin of the area term in the non-singlet sectors of the theory.}

\begin{document}

\maketitle

\section{Introduction}

In many models of quantum gravity, spacetime itself can be understood as emerging from entanglement in a dual quantum mechanical formulation. Perhaps the most prominent example of such a phenomenon is found in the AdS/CFT correspondence, where the Ryu-Takayanagi (RT) formula \cite{Ryu:2006bv} connects entanglement entropy in the boundary CFT to spacetime geometry in the AdS space. However, the spatial entanglement that is considered in the RT formula on its own cannot explain all features of the AdS geometry. In general, it is believed that internal degrees of freedom are needed to describe regions deep in the bulk \cite{Balasubramanian:2014sra,Balasubramanian:2016xho,Balasubramanian:2018ajb,Erdmenger:2019lzr,Gerbershagen:2021gvc,Craps:2022pke,Gerbershagen:2024ram}, locality on sub-AdS length scales \cite{Balasubramanian:2005qu,Heemskerk:2009pn,Balasubramanian:2014sra,Ghosh:2024myq} or the geometry of compact internal spaces in top-down AdS/CFT setups \cite{Mollabashi:2014qfa,Karch:2014pma,Das:2022njy,Bohra:2024wrq}. These internal degrees of freedom oftentimes organize themselves in matrix form. An extreme case of the emergence of spacetime from entanglement is then found in matrix quantum mechanics (MQM), where there is no spacetime to begin with and all of the geometry should emerge from entanglement between internal matrix degrees of freedom \cite{Mazenc:2019ety,Das:2020jhy,Das:2020xoa,Hampapura:2020hfg,Gautam:2022akq} (see \cite{Fliss:2025omb} for a review).

The focus of this paper lies on the MQM associated to the $c=1$ two-dimensional string theory.\footnote{Reviews on this topic include \cite{Klebanov:1991qa,Martinec:1991kn,Ginsparg:1993is,Polchinski:1994mb,Martinec:2004td}.} Briefly summarized, this model contains the following ingredients. The worldsheet formulation is given by a single bosonic worldsheet field coupled to Liouville theory, giving a critical string theory in a two-dimensional target spacetime spanned by the boson and the Liouville field. The low-energy effective field theory is given by 2d dilaton gravity coupled to a scalar field. The perturbative string theory admits a reformulation in terms of a double-scaled MQM with an inverted harmonic oscillator potential. The description of the 2d target spacetime background in this model in terms of MQM variables is thus an interesting example of an emergent spacetime which lies outside of the usual AdS/CFT paradigm.

Entanglement in the $c=1$ model has been examined in \cite{Das:1995vj,Hartnoll:2015fca}. In this work, we focus on a particular puzzle related to these results which has not been emphasized before. In general, one expects a well-defined measure of the entanglement of a subregion $\cR$ in a gravitational theory to be the generalized entropy, given as the sum of a gravitational piece proportional to the area of the subregion boundary and the entanglement entropy of matter fields in the subregion $\cR$  \cite{Susskind:1994sm},
\begin{equation}
    S_\text{gen}(\cR) = \frac{A_{\partial \cR}}{4G_N} + S_\text{mat}(\cR).
    \label{eq:generalized-entropy}
\end{equation}
In 2d dilaton gravity models, such as the one forming the low-energy effective field theory of the $c=1$ string theory, the expectation is that the area of the subregion boundary is replaced by the value of the dilaton at the subregion boundary (which is just a collection of points in 2d), see e.g.~\cite{Almheiri:2019psf}.\footnote{For ease of terminology, we will sometimes refer to such area-like terms as just area terms in the following.}
However, such an area-like term was entirely absent in the previous studies of entanglement in the $c=1$ matrix model \cite{Das:1995vj,Hartnoll:2015fca}.
Rather, the results of \cite{Das:1995vj,Hartnoll:2015fca} can be interpreted as only yielding the contribution to the generalized entropy from the matter field, i.e.~the tachyon in the $c=1$ model, as was noted in \cite{Mazenc:2019ety}.
In particular, ref.~\cite{Hartnoll:2015fca} finds
\begin{equation}
    S = \frac{1}{3}\log\frac{\mu (x_2-x_1)}{\sqrt{g_s(x_1)g_s(x_2)}} + \dots \, ,
    \label{eq:Hartnoll-Mazenc-result}
\end{equation}
where $x_1,x_2$ are the endpoints of the interval $\cR$, and dots represent extra terms for which the computation was not fully reliable but that did not correspond to an area term. Eq.~\eqref{eq:Hartnoll-Mazenc-result} is the same as the entanglement entropy $S = \frac{1}{3}\log((x_2-x_1)/\varepsilon)$ for a free scalar in two dimensions with the UV cutoff $\varepsilon = \sqrt{g_s(x_1)g_s(x_2)}/\mu$ set by the string coupling, establishing finiteness of the entanglement entropy in 2d string theory.

The apparent mismatch between \eqref{eq:generalized-entropy} and \eqref{eq:Hartnoll-Mazenc-result} leads us to the central question motivating this work:

\begin{center}
\emph{Is there an area-like term for the generalized entropy in the $c=1$ 2d string theory, and if so, where to find it in the dual matrix quantum mechanics?}
\end{center}

After a brief review of the $c=1$ model and its dual MQM formulation in \secref{sec:Review}, we address the following points related to the above question.

In \secref{sec:EntanglementTargetSpace}, we present a number of arguments that -- although falling short of a fully conclusive proof -- support the existence of an area term in the $c=1$ model. This section serves as a point of reference to collect existing discussions of area terms and extend them to arbitrary regions in the 2d dilaton gravity setup of interest to us. In particular, we consider the following points. First, we discuss how the argument of Susskind and Uglum \cite{Susskind:1994sm} showing cancellation of UV divergences between the area and matter terms of the generalized entropy does not work in the same way in two dimensions as UV divergences are cancelled by a topological piece instead. We then study the gravitational thermodynamics in the low-energy effective field theory as well as an algebraic approach based on introducing additional observer degrees of freedom. Both of these computations indicate that an area term should be present, thus suggesting that the question we are asking is well-defined.

One noteworthy point is that the aforementioned arguments address the existence of an area term for an arbitrary subregion in the spacetime background corresponding to the ground state of the double-scaled MQM. This question is distinct from asking whether an area-like term exists in the entropy of a black hole in the 2d dilaton gravity of the $c=1$ model. While the latter question has been studied in the past \cite{Gibbons:1992rh,Nappi:1992as} and can be addressed for general 2d dilaton gravity models \cite{Grumiller:2007ju}, the situation that is of interest to us is more subtle. For this reason, more sophisticated arguments are necessary; collecting such arguments is the purpose of section \ref{sec:EntanglementTargetSpace}.

The apparent mismatch between entanglement in the MQM and the arguments in the low-energy EFT begs the question as to where to find the area term in the matrix quantum mechanics dual to the $c=1$ model.
Asking in another way: Between which degrees of freedom in the MQM should one compute the entanglement entropy in order to reproduce the generalized entropy associated to an interval in the target spacetime where the EFT is defined?

There are two natural places to look for an area term. The first is a nonlocal transformation connecting the taychon field of the EFT with a scalar describing the collective field theory of excitations on top of the Fermi level in the singlet sector of the MQM (see \secref{sec:Review} for details). This transformation arises from additional ``leg-pole'' factors that are needed to match worldsheet string amplitudes with corresponding calculations in the MQM \cite{Klebanov:1991qa,Klebanov:1991ai,Natsuume:1994sp}. It has been shown that the nonlocal transformation encodes gravitational scattering effects for the tachyon field \cite{Natsuume:1994sp}. Previous computations of entanglement entropy in the $c=1$ model \cite{Das:1995vj,Hartnoll:2015fca} were restricted to studying entanglement of an interval in the space of eigenvalues of the singlet sector of the MQM. The effect of the nonlocal transformation is thus not visible in these computations. Given that this transformation encodes gravitational effects in tachyon scattering, it is natural to expect that it will also encode gravitational effects in the entanglement entropy. As shown in \secref{sec:entanglement-MM}, we indeed find this to be true. However, for an interval far away from the tachyon wall these effects are small perturbative corrections to the matter entropy instead of the large deviations expected from an area term. We thus conclude that the nonlocal transformation does not generate an area term in the entanglement entropy.

Stepping back from the narrow question of whether the inclusion of the nonlocal transformation generates an area term, one might wonder what the computation tells us about the emergence of spacetime from entanglement. In that direction, the somewhat surprising conclusion is that even though important gravitational scattering effects are encoded in the nonlocal transformation, its effect on the entanglement structure is quite modest. It neither affects finiteness of the entanglement entropy\footnote{The finiteness is not visible in our computation, but this is solely due to the approximations used in order to make the problem tractable. It is clear that without those approximations, we would obtain a finite answer.} nor drastically changes the actual numerical value of the entanglement entropy away from the strongly coupled region. Loosely speaking, even though degrees of freedom that are localized within an interval are spread out over the entire spatial direction by the nonlocal transformation, the contribution from degrees of freedom that are transported far away from the interval is small enough to keep the entanglement structure largely intact.

The second natural place where an area term could appear is the contribution of non-singlets to the entanglement entropy. Previous discussions as well as that of \secref{sec:entanglement-MM} only consider entanglement between degrees of freedom in the singlet sector of the MQM. While this sector correctly describes the physics of small excitations of the ground state \cite{Klebanov:1991qa}, larger effects such as black hole geometries in the EFT are not described by the singlets alone \cite{Karczmarek:2004bw}. Therefore, given that the area term does not arise from the nonlocal transformation, it is natural to expect it to be encoded in the contribution of non-singlets instead. While confirming this expectation is out of the scope of this publication, we offer some comments on this issue in \secref{sec:non-singlets}. Naively one might expect the entanglement entropy in the ground state, which is supported entirely on the singlet degrees of freedom, not to have any non-singlet contributions due to the linearity of the partial trace. However, by choosing a suitable bipartition of the matrix degrees of freedom, non-singlets will contribute to the entanglement entropy as we explain by analogy with the simpler setting of distinguishable versus indistinguishable particles reviewed in appendix~\ref{app:EntanglementNonInvariant}. Therefore, it is possible that non-singlet contributions determine the area term in the generalized entropy from the MQM, although the exact nature of the bipartition to choose for the non-singlet degrees of freedom is unclear at the moment.

We conclude in \secref{sec:subsystems-and-subregions} by making some cautionary comments on the above analysis. Implicitly assumed in the discussion was that a generic spatial subregion in the low-energy effective field theory is described by a well-defined subsystem in the MQM. This assumption is necessary for the question about the existence of an area term in the MQM to be well-posed in the first place. However, recent arguments, which introduce the notion of generalized entanglement wedges as the only subregions associated to well-defined subsystems, cast doubt on this assumption \cite{Bousso:2022hlz,Bousso:2023sya}. While it is an open question if such arguments apply to the $c=1$ model, we discuss the implication a positive answer to this question would have for the discussion about area terms.

\section{Review of the $c=1$ theory}
\label{sec:Review}

In this section, we will briefly review the target space and matrix quantum mechanics descriptions of the $c = 1$ model, primarily as a way to establish our conventions. A third description is provided by the worldsheet perspective, which we will not use in this work; we will briefly comment on it in footnote~\ref{footnote:Worldsheet}.

\subsection{Target space}

From the target space perspective, we have bosonic strings propagating on a two-dimen\-sional background. This background has to satisfy the equations of motion of the low-energy effective theory, which is a dilaton gravity model with an extra massless field traditionally called the tachyon (since it comes from the same kind of bosonic string excitations that produce tachyonic modes in higher-dimensions). We split it as
\begin{equation}
\label{Review:TargetSpaceAction}
\cI = a_1 \cI_{0} + \cI_T \, ,
\end{equation}
with $a_1$ a constant to be fixed below, $\cI_0$ the dilaton gravity piece
\begin{equation}
\label{Review:DilatonGravityAction}
\cI_0 = \frac{1}{2} \int \rd t \rd\phi \, \sqrt{-G} e^{-2 \Phi} \left( R + \frac{16}{\alpha'} + 4 (\partial \Phi)^2 \right) \, ,
\end{equation}
and $\cI_T$ the tachyon part
\begin{equation}
\label{Review:TachyonAction}
\cI_T = \frac{1}{2} \int \rd t \rd\phi \, \sqrt{-G} e^{-2 \Phi} \left( - (\partial T)^2 + \frac{4}{\alpha'} T^2 - 2 V(T) \right) \, .
\end{equation}
The potential $V(T)$ contains tachyon self-interactions that do not play any role in our discussion, so we leave it unspecified.

We focus on a specific linear-dilaton solution of this theory. For the metric and dilaton, this is
\begin{equation}
\label{Review:MetricDilatonBackground}
G^{(0)}_{\mu \nu} \rd x^{\mu} \rd x^{\nu} = \eta_{\mu \nu} \rd x^{\mu} \rd x^{\nu} = - \rd t^2 + \rd \phi^2 \, , \qquad \Phi^{(0)} = \frac{2 \phi}{\sqrt{\alpha'}} \equiv \frac{2 \phi}{\ell_s} \, .
\end{equation}
Notice that the effective string coupling behaves as $e^{2 \phi / \ell_s}$, becoming exponentially weak as $\phi \to - \infty$. In that weakly coupled region, we can find the leading order tachyon profile from its linearized equation of motion
\begin{equation}
\label{Review:TachyonBackground}
T^{(0)} = \left( b_1 \frac{\phi}{\ell_s} + b_2 \right) e^{2 \phi / \ell_s} \, ,
\end{equation}
with the values of the constants $b_1$ and $b_2$ given below. The background $(G^{(0)}, \Phi^{(0)}, T^{(0)})$ solves each of the equations of motion to leading non-trivial order in powers of $T^{(0)}$ -- see \eqref{CovariantPhaseSpace:EquationsOfMotion} for the explicit form of these equations. To higher orders, we need to take into account both corrections in the fields and in the effective action, including tachyon self-interactions \cite{Polchinski:1990mf}. Our discussion will be entirely within the weakly coupled region $\phi / \ell_s \to - \infty$, so having the leading solution is enough.\footnote{\label{footnote:Worldsheet}From a worldsheet perspective, this background comes from a known CFT: a $c=1$ free scalar providing the timelike dimension together with the $c=25$ Liouville CFT (plus the diffeomorphism ghosts). See \cite{Klebanov:1991qa,Martinec:1991kn,Ginsparg:1993is,Polchinski:1994mb} for classic reviews, \cite{BalthazarPhDThesis} for an up to date approach, and \cite{Martinec:1991kn,Ginsparg:1993is} for comments on higher-order corrections to the background.}

In this background, the relevant physical process is the scattering of closed strings that come in from $\phi \to - \infty$, interact near the exponential tachyon wall and reflect back to the asymptotic region. The matrix quantum mechanical model discussed below precisely captures two-dimensional string theory on this background. In fact, matching the target space and matrix model scattering amplitudes fixes the constants above to $a_1 = 1/2$, $b_1= \mu/\sqrt{4 \pi}$ and $b_2 = -b_1 (1+4\Gamma'(1)-\log \mu)$ \cite{Natsuume:1994sp}, with $\mu$ a matrix model parameter defined in \eqref{Review:DoubleScalingLimit}.\footnote{Similar computations fix the first interaction in the tachyon potential to be $V(T) = -2 \sqrt{2} T^3/(3\alpha')$, but this will not be relevant for our discussion.}

\subsection{The $c=1$ matrix quantum mechanics}

In order to introduce the $c=1$ matrix quantum mechanical model, consider a single $N \times N$ Hermitian matrix $M$ whose dynamics is governed by the action
\begin{equation}
\label{Review:MQMAction}
\cI = \beta N \int \rd t\, \Tr\left( \frac{1}{2} \dot{M}^2(t) - V(M(t)) \right)
\end{equation}
with a potential that has a maximum at the origin, e.g.,\footnote{In the remainder of the paper, we are working in units where $\alpha'=\ell_s=1$ unless indicated otherwise.}
\begin{equation}
V(M) = \frac{1}{4} M^2(M^2-2) \, .
\end{equation}
The low-energy physics involves singlet states, invariant under the $SU(N)$ symmetry. These singlet states are characterized by the $N$ eigenvalues $\lambda_1, \dots, \lambda_N$ of the matrix, which behave as $N$ independent fermions moving under the influence of the potential $V(\lambda)$ \cite{Brezin:1977sv}.

In the large-$N$ limit, this system is well known to have a critical behavior as $\beta$ approaches (from above) some critical value $\beta_c$ which depends on the specifics of the potential -- for the previous example, $\beta_c = 3 \pi /2$ \cite{Klebanov:1991qa,Polchinski:1994mb}. This can be understood as follows. In terms of the fermions, we are occupying $N$ states in the valley to the left of the maximum at $\lambda = 0$, up to some Fermi level $- E_F < 0$.\footnote{At finite $N$, the states should spread to both sides of the finite barrier. However, we will be interested in a large-$N$ limit, where states can be localized to just one side.} $\beta N$ is playing the role of an effective $\hbar^{-1}$, so the spacing between levels is of order $(\beta N)^{-1}$. Having $N$ occupied levels, the range of occupied energies scales as $\beta^{-1}$, and thus as we reduce $\beta$ there is a point at which the Fermi level (which is determined by $\beta$) crosses the maximum of the potential barrier at zero energy. Approaching this point, i.e., taking $\beta \to \beta_c$ such that $E_F \to 0^{+}$, we get the critical behavior.

The critical point can be approached in a double-scaling limit, and this is the regime in which the matrix system provides a dual description of two-dimensional string theory in the background discussed in the previous section. This is done by taking
\begin{equation}
\label{Review:DoubleScalingLimit}
N \to \infty \, ,\qquad \beta \to \beta_c \, , \qquad \beta N E_F(\beta) \equiv \mu \;\; \text{fixed} \, .
\end{equation}
Note that $E_F$ here refers to an energy normalized as $p^2/2 + V(\lambda)$, i.e., we have absorbed $\beta N$ into an effective $\hbar^{-1}$ as described above. Therefore, the $N \to \infty$ limit makes the theory classical and we can describe it in terms of excitations on top of a Fermi sea. Rescaling the eigenvalue coordinate as $x = (\beta N)^{1/2} \lambda$, the surface of the Fermi sea is
\begin{equation}
\frac{p^2}{2 \beta N} + \frac{x^2}{4 \beta N} \left(\frac{x^2}{\beta N} - 2 \right) = - E_F \, .
\end{equation}
As $\beta N \to \infty$, interactions in the original potential can be ignored: the double scaling limit is focusing around the maximum of the potential in such a way that higher-order terms become negligible. This is a general lesson, valid irrespectively of the particular form of $V(\lambda)$ we started with. The Fermi sea is then the curve in phase space
\begin{equation}
\frac{p^2}{2} - \frac{x^2}{2} = - \mu \, .
\end{equation}

This equation can be seen as the definition of the parameter $\mu$ in the double-scaling limit: it provides the value of the equilibrium Fermi surface. In fact, analyzing the perturbative expansion of the double-scaled matrix quantum mechanics, one finds that it reproduces the topological expansion of a string theory with string coupling $g_s \sim \mu^{-1}$ \cite{Klebanov:1991qa,Polchinski:1994mb}. From the string theory perspective, $\mu$ is thus setting the coupling strength.\footnote{As usual, this statement requires a convention to fix the dilaton shift symmetry. In the way we wrote the background \eqref{Review:MetricDilatonBackground} and \eqref{Review:TachyonBackground}, it really means that when $e^{2 \Phi} \sim 1$, the tachyon wall has strength $b_1, b_2 \sim \mu$.} 

In order to make contact with the target space picture, a reformulation of the fermionic theory in terms of a collective field capturing excitations on top of the Fermi surface is convenient \cite{Das:1990kaa,Polchinski:1991uq}.
In second quantization, the Hamiltonian of the fermionic theory is given by
\begin{equation}
    H = \int \rd x \left[\frac{1}{2}\frac{\partial \psi^\dagger}{\partial x}\frac{\partial\psi}{\partial x} + \left(\mu-\frac{x^2}{2}\right)\psi^\dagger\psi\right].
    \label{eq:fermionic-Hamiltonian}
\end{equation}
Applying the standard bosonization prescription to the fermionic theory gives a Hamiltonian \cite{Klebanov:1991qa}
\begin{equation}
  \begin{aligned}
    H = \frac{1}{2}\int \rd \tau \biggl[&\bar\Pi^2 - (\partial_\tau \bar S)^2 - \frac{\sqrt\pi}{2\mu\sinh^2\tau}\left(\bar\Pi^2\partial_\tau \bar S+\frac{1}{3}(\partial_\tau \bar S)^3\right)\\
                               &+ \frac{\partial_\tau \bar S}{8\mu\sqrt\pi}\left(\frac{1}{3\sinh^2\tau} + \frac{1}{\sinh^4\tau}\right)\biggr] \, ,
  \end{aligned}
  \label{eq:bosonic-Hamiltonian}
\end{equation}
where $\bar S$ is the collective field describing excitations on top of the Fermi sea, $\bar\Pi$ its canonically conjugate momentum, and $\tau$ the so-called `time-of-flight' coordinate, defined by
\begin{equation}
    \frac{\rd x}{\rd\tau} = v(x) = \sqrt{x^2-2\mu} \, .
    \label{eq:definition-time-of-flight-variable}
\end{equation}
Here, $v(x)$ is the velocity of a particle moving along a classical trajectory on the Fermi sea. Note that while bosonization in a relativistic setting relates free fermions to free bosons, in the non-relativistic theory that we are considering the bosonic Hamiltonian contains cubic interaction terms.

The collective field description is very closely related to the target space picture, in which the only propagating degree of freedom is the massless tachyon. There is however one important and non-trivial ingredient. Expand in modes the collective field
\begin{equation}
  \bar{S}(t,\tau) = \frac{1}{i\sqrt{2}} \int_{-\infty}^\infty \frac{\rd\omega}{2\pi\omega}\left(\bar{\alpha}_+(\omega) e^{i\omega(t-\tau)}\left(\frac{2}{\mu}\right)^{-i\omega/2} + \bar{\alpha}_-(\omega)e^{i\omega(t+\tau)}\left(\frac{2}{\mu}\right)^{i\omega/2}\right) \, ,
  \label{eq:Fourier-expansion-collective-field}
\end{equation}
with $\bar\alpha_\pm(\omega)$ the momentum space creation/annihilation operators. In the target space theory, we define a rescaled tachyon field $S(t,\phi) = e^{-2\phi}T(t,\phi)$, which we also expand in modes as
\begin{equation}
  S(t,\phi) = \frac{1}{i\sqrt{2}} \int_{-\infty}^\infty \frac{\rd\omega}{2\pi\omega}\left(\alpha_+(\omega) e^{i\omega(t-\phi)}\left(\frac{2}{\mu}\right)^{-i\omega/2} + \alpha_-(\omega)e^{i\omega(t+\phi)}\left(\frac{2}{\mu}\right)^{i\omega/2}\right) \, .
\end{equation}
The dynamics of $\bar{S}$ is fixed by the fermionic theory derived from the matrix quantum mechanics, that of $S$ comes from the target space effective action. Computing scattering amplitudes in both theories, the tree-level S-matrices agree, provided the momentum space creation and annihilation operators are related via the `leg-pole' factors \cite{DiFrancesco:1991daf,DiFrancesco:1991ocm,Polyakov:1991qx}
\begin{equation}
  \alpha_\pm(\omega) = \left(\frac{\pi}{2}\right)^{\pm i \omega/4} \frac{\Gamma(\pm i \omega)}{\Gamma(\mp i \omega)} \bar{\alpha}_\pm(\omega) \, .
  \label{eq:leg-pole}
\end{equation}

The leg-pole factors amount to a local transformation in momentum space, but they induce a nonlocal transformation between the fields in real space \cite{Natsuume:1994sp},
\begin{equation}
  S(t,\phi) = \int_{-\infty}^\infty \rd\tau K(\phi - \tau) \bar S(t,\tau)\, ,
\label{eq:nonlocal-transformation}
\end{equation}
where the integration kernel is given by
\begin{equation}
    K(\tau) = \int_{-\infty}^\infty \frac{\rd\omega}{2\pi}e^{i\omega\tau} \left(\frac{\pi}{2}\right)^{-i\omega/4}\frac{\Gamma(-i\omega)}{\Gamma(i\omega)}= - \frac{z}{2}J_1(z) \, , \qquad z = 2(2/\pi)^{1/8}e^{\tau/2} \, .
\label{eq:integration-kernel-nonlocal-transformation}
\end{equation}
This transformation generates gravitational scattering in the target space effective theory.
In the matrix model, a fermionic excitation will propagate undisturbed in the harmonic oscillator potential due to the non-interacting nature of the theory, so no scattering will occur in this description.
However, in the target space picture a tachyonic excitation will produce a gravitational field which can then scatter other incoming tachyons.
The effect of this scattering is encoded in the nonlocal transformation \eqref{eq:nonlocal-transformation} \cite{Natsuume:1994sp}.

\section{Area terms from the target space effective field theory}
\label{sec:EntanglementTargetSpace}

In the present section, we will start from the target space effective theory \eqref{Review:TargetSpaceAction} and provide arguments that support the existence of an area-like term (i.e., a term that depends exponentially on the dilaton value) in the entropy of a target space subregion. For simplicity and concreteness, we will mostly focus on a single interval, but the results generalize to other subregions. We start in section \ref{subsec:UVFiniteness} by reviewing an argument by Susskind and Uglum \cite{Susskind:1994sm} which shows that the combination of gravitational and matter terms in the generalized entropy has better UV properties than its constituents. This argument presents some known subtleties in the context of 2d dilaton gravity models; we discuss these in detail and conclude that by itself the Susskind-Uglum construction does not require (but does not rule out either) a dilaton dependent area term in the generalized entropy of a target space subregion. Section \ref{subsec:GravitationalThermodynamics} then collects other arguments that support more clearly the existence of an area term: one based on the thermodynamic properties of causal diamonds \cite{Jacobson:2015hqa,Jacobson:2018ahi}, and another one relying on the algebraic entropy associated to a bulk observer which can only access a spacetime subregion \cite{Jensen:2023yxy}. While these constructions were known in higher-dimensional contexts, we implement them in the 2d target space effective theory and show that their conclusions still hold. We finally assess in section \ref{subsec:CommentsArguments} what each of the previous arguments really allows us to conclude about the presence of a dilaton dependent area term in the generalized entropy of a target space subregion.

\subsection{UV finiteness of the generalized entropy}
\label{subsec:UVFiniteness}

A classic argument by Susskind and Uglum \cite{Susskind:1994sm} has been taken for a long time as an indication that, in gravitational theories, the generalized entropy is a quantity with better UV properties than its constituents taken separately (i.e., the gravitational area-like contribution and the matter entanglement entropy) -- see \cite{Bousso:2015mna} for a detailed review of this claim. In more than two dimensions, the generalized entropy of a codimension-$1$ spatial region $\cR$ is given by
\begin{equation}
\label{EntanglementTargetSpace:GeneralizedEntropyHigherD}
S_{\rm gen}(\cR) = \frac{A_{\partial \cR}}{4 G_N} + S_{\rm mat}(\cR) \, ,
\end{equation}
where $A_{\partial\cR}$ is the area of the boundary of $\cR$, and $S_{\rm mat}(\cR)$ is the von Neumann entropy of the reduced state of the fields in (the domain of dependece of) $\cR$. We are implicitly assuming that the theory is General Relativity minimally coupled to some matter fields; more general theories (e.g., with higher-curvature terms) receive extra contributions in the gravitational entropy. In the two-dimensional theory \eqref{Review:TargetSpaceAction}, the analogous expression for the generalized entropy is
\begin{equation}
\label{EntanglementTargetSpace:GeneralizeEntropy2d}
S_{\rm gen}(\cR) = \left. 2 \pi a_1 e^{-2 \Phi} \right|_{\partial \cR} + S_{\rm mat}(\cR) \, ,
\end{equation}
where now $\partial \cR$ is the set of points that bounds the one-dimensional region $\cR$, and the tachyon is the only field contributing to the matter entropy. The form of the gravitational term, including its particular coefficient and the dilaton dependence, will be justified in the next subsection and appendix~\ref{app:CovariantPhaseSpace} via a covariant phase space analysis. For now, we just take it as given, noting that this form of the gravitational contribution in dilaton gravities has been extensively discussed in the literature \cite{Gibbons:1992rh,Nappi:1992as,Grumiller:2007ju}.

The essence of the Susskind-Uglum argument is the following. In the generalized entropy \eqref{EntanglementTargetSpace:GeneralizedEntropyHigherD}, both the bare gravitational coupling which enters the area term and the entropy of the bulk matter fields are UV-divergent quantities. If these divergences are separately computed, in most situations they cancel, leaving a finite generalized entropy written in terms of the renormalized gravitational coupling and the bulk matter entropy with UV divergences subtracted. Here we briefly discuss how much of this argument extends to the theory \eqref{Review:TargetSpaceAction}. As already noted in \cite{Susskind:1994sm}, this case has some particular features which are different from those of standard higher-dimensional gravitational models.

The first step is to compute the effective action that arises after integrating out the tachyon, since this determines the renormalization of the gravitational coupling. We work in Euclidean signature in this section. Leaving the details to appendix~\ref{app:GeneralizedEntropy}, the effective action after integrating out the tachyon in \eqref{Review:TachyonAction} (ignoring self-interactions) contains a UV-divergent piece proportional to the integral of the two-dimensional conformal anomaly \cite{Mukhanov:1994ax,Kummer:1997jr}. Within it, the curvature dependent term is
\begin{equation}
\label{EntanglementTargetSpace:EffectiveActionCurvature}
W \supset \frac{1}{48 \pi} \int \rd^2 x \sqrt{G} \, \log \left( \epsilon^2 \right) R \, ,
\end{equation}
with $\epsilon$ a length cutoff. Notice that this does not have a dilaton factor $e^{-2 \Phi}$, and thus it does not renormalize $a_1$ in \eqref{Review:TargetSpaceAction}. Instead, we can imagine we add an extra piece to the effective action of the form
\begin{equation}
\label{EntanglementTargetSpace:TopologicalTerm}
- b_B\int \rd^2 x \sqrt{G} \, R \, ,
\end{equation}
with $b_B$ the bare coupling. Note that this term does not affect the dynamics, since the integral of $R$ in two-dimensions just gives the Euler characteristic of the manifold. The previous result shows that the one-loop renormalized value is
\begin{equation}
b = b_B - \frac{1}{48 \pi} \log (\epsilon^2) \, .
\end{equation}

The addition of the term \eqref{EntanglementTargetSpace:TopologicalTerm} to the action produces an extra term in the gravitational entropy. Essentially the same logic that led to \eqref{EntanglementTargetSpace:GeneralizeEntropy2d} (or the computation in appendix~\ref{app:CovariantPhaseSpace}) gives as the new candidate generalized entropy of an interval $(\phi_L, \phi_R)$
\begin{equation}
S_{\rm gen} = 8 \pi b_B + 2 \pi a_1 e^{-2 \Phi(\phi_L)} + 2 \pi a_1 e^{-2 \Phi(\phi_R)} + S_{\rm mat} \, ,
\end{equation}
with $S_{\rm mat}$ the von Neumann entropy of the reduced state of the tachyon field in the interval. Note that the first term combines the contribution from both boundaries of the interval ($4 \pi b_B$ from each of them). The tachyon is a massless scalar propagating in the flat linear-dilaton background, so its entropy on an interval is known to be \cite{Calabrese:2004eu}
\begin{equation}
S_{\rm mat} = \frac{1}{3} \log \left( \frac{\phi_R - \phi_L}{\epsilon} \right) \, .
\end{equation}
It can be readily checked that this UV divergence provides the necessary term to transform $b_B$ into its renormalized value.

As already found in \cite{Susskind:1994sm}, in two-dimensional dilaton gravity models the matter fields renormalize the topological term of the action, rather than the $e^{-2 \Phi} R$ piece. From this viewpoint, the status of the Susskind-Uglum argument is different than in higher-dimensions: the existence of a piece proportional to $e^{-2 \Phi}$ in the generalized entropy is not excluded, but also not required to get a quantity with good UV behavior. We leave a critical assessment of this fact to section \ref{subsec:CommentsArguments}, and turn now to other arguments that signal the presence of an area-like term.

\subsection{Gravitational thermodynamics, algebras, and generalized entropy}
\label{subsec:GravitationalThermodynamics}

It is well-known that the target space effective theory \eqref{Review:TargetSpaceAction} supports black hole solutions \cite{Mandal:1991tz,Witten:1991yr,Dijkgraaf:1991ba}. In the presence of horizons, a convenient way to derive the form of the black hole entropy and the first law it satisfies is via covariant phase space methods \cite{Wald:1993nt,Iyer:1994ys}. Even though our focus is on the linear dilaton background, in which no black hole is present, it will be useful to understand first how an area-like term appears in the black hole entropy for the two-dimensional theory \eqref{Review:TargetSpaceAction}. In this section, we will frequently use technical results collected in appendix~\ref{app:CovariantPhaseSpace}, where we review the essential ingredients of the covariant phase space formalism, applied to the theory \eqref{Review:TargetSpaceAction}. 

Consider a black hole solution with a bifurcate horizon in the theory \eqref{Review:TargetSpaceAction}, and let $\cR$ be a time slice extending from the bifurcation surface $\cB$ to spatial infinity, which we denote by $\cB_{\infty}$. Let $\xi$ be the Killing vector generating the bifurcate horizon, which at the bifurcation surface satisfies \cite{Wald:1993nt,Wald:1984rg}
\begin{equation}
\label{EntanglementTargetSpace:HorizonKilling}
\left. \xi^{\mu} \right|_{\cB} = 0 \, , \qquad \left. \nabla_{\mu} \xi_{\nu} \right|_{\cB} = \kappa \, \epsilon_{\mu \nu} \, ,
\end{equation}
where $\kappa$ is the surface gravity and $\epsilon_{\mu \nu}$ is the volume form of the spacetime.\footnote{In higher-dimensions, the role of $\epsilon_{\mu \nu}$ is played by the binormal form to the bifurcation surface. In two-dimensions, the binormal coincides with the spacetime volume form.} Given that $\xi$ is a Killing vector, the analysis of appendix~\ref{app:CovariantPhaseSpace} leads to the identity \eqref{CovariantPhaseSpace:FirstLaw},
\begin{equation}
\label{EntanglementTargetSpace:FirstLawAbstract}
0 = \left. \left( \delta Q_{\xi} - \xi \cdot \Theta \right) \right|_{\cB_{\infty}} - \left. \left( \delta Q_{\xi} - \xi \cdot \Theta \right) \right|_{\cB} \, ,
\end{equation}
where $Q_{\xi}$ is given by \eqref{CovariantPhaseSpace:NoetherCharge},
\begin{equation}
\label{EntanglementTargetSpace:NoetherCharge}
Q_{\xi} = - \frac{e^{-2 \Phi}}{2} a_1 \epsilon^{\mu \nu} \left(\nabla_{\mu} \xi_{\nu} - 4 \xi_{\mu} \nabla_{\nu} \Phi \right) \, ,
\end{equation}
and $\xi \cdot \Theta$ follows from \eqref{CovariantPhaseSpace:Theta},
\begin{equation}
\xi \cdot \Theta = a_1 e^{-2 \Phi} \left[ 4 \delta \Phi \nabla^{\rho} \Phi + g^{\mu [\rho} \nabla^{\nu]} \delta g_{\mu \nu} + 2\left( g^{\mu [\rho} \nabla^{\nu]} \Phi \right) \delta g_{\mu \nu} - \frac{1}{a_1} \delta T \nabla^{\rho} T \right] \xi^{\sigma} \epsilon_{\rho \sigma} \, .
\end{equation}
In the previous expressions, $\delta$ is to be interpreted as a first-order variation of a solution to a nearby one (e.g., a small change in the parameters defining the black hole).

We can use the properties of the Killing field at $\cB$ collected in \eqref{EntanglementTargetSpace:HorizonKilling} to rewrite \eqref{EntanglementTargetSpace:FirstLawAbstract} as
\begin{equation}
\left. \delta Q_{\xi} \right|_{\cB} = \left. a_1 \kappa \, \delta e^{-2 \Phi} \right|_{\cB} = \left. \left( \delta Q_{\xi} - \xi \cdot \Theta \right) \right|_{\cB_{\infty}} \, .
\end{equation}
This is the classical first law for the black hole horizon. The right hand side, evaluated at the asymptotic boundary, will give the contribution from the charges, while the horizon piece gives the entropy variation. Identifying the Hawking temperature from the surface gravity as $T_H = \kappa/(2 \pi)$, the area-like form of the entropy is
\begin{equation}
S_{\rm BH} = \left. 2 \pi a_1 e^{-2 \Phi} \right|_{\cB} \, .
\end{equation}
It is worth noting that this result coincides with computations based on the Euclidean approach to black hole thermodynamics \cite{Gibbons:1992rh,Nappi:1992as}.

Given that this area term exists for a black hole horizon, it is not unreasonable to expect that such a gravitational contribution to the entropy appears at the boundaries of any spatial subregion. Still, it is unsatisfactory to rely on a result valid in the presence of a black hole horizon to argue for the existence of an area-like term in a generic subregion when no black hole is present. Let us consider then the situation we are interested in, namely the causal diamond which is the domain of dependence of an interval $(\phi_L, \phi_R)$ in flat two-dimensional Minkowski space (we put the interval at $t=0$). In the linear dilaton background \eqref{Review:MetricDilatonBackground}, we imagine that this interval is taken at large and negative values of $\phi_L$ and $\phi_R$, so that it is well within the region in which a semiclassical spacetime exists.

Introduce a new coordinate $y$ centered at the midpoint of the interval, which becomes now $y \in (-L/2, L/2)$. The causal diamond has an associated conformal Killing vector explicitly given by \cite{Jacobson:2015hqa}
\begin{equation}
\label{EntanglementTargetSpace:CausalDiamondCKV}
\zeta = \frac{(L/2)^2 -t^2-y^2}{L} \partial_t - \frac{2 t y}{L} \partial_y \, .
\end{equation}
The flow of this vector maps the diamond to itself, $\zeta$ vanishes at the boundaries of the interval, and
\begin{equation}
\left. \nabla_{\mu} \zeta_{\nu} \right|_{t=0, y=L/2} = - \epsilon_{\mu \nu} \, , \quad \left. \nabla_{\mu} \zeta_{\nu} \right|_{t=0, y=-L/2} = \epsilon_{\mu \nu} \, .
\end{equation}
Given that $\zeta$ vanishes at the boundaries of the interval $\cR = (-L/2, L/2)$, we can obtain the relation \eqref{CovariantPhaseSpace:HamiltonianIntegrated},
\begin{equation}
\label{EntanglementTargetSpace:HamiltonianSubregion}
H_{\zeta}^{\cR} = \int_{\cR} F_{\zeta} + \int_{\partial \cR} Q_{\zeta} = \int_{\cR} F_{\zeta} - a_1 e^{-2 \Phi(\phi_R)} - a_1 e^{-2 \Phi(\phi_L)} \, .
\end{equation}
Here, $H_{\zeta}^{\cR}$ is a Hamiltonian function that generates the diffeomorphism transformation along $\zeta$ for the fields within $\cR$, $Q_{\zeta}$ has been evaluated using \eqref{EntanglementTargetSpace:NoetherCharge}, and $F_{\zeta}$ is a one-form given by \eqref{CovariantPhaseSpace:ConstraintGenerator}:
\begin{equation}
F_{\zeta} = - a_1 e^{-2\Phi} \zeta_{\lambda} \cE^{\lambda \mu} \epsilon_{\mu \nu} \rd x^{\nu} \, ,
\end{equation}
with $\cE^{\lambda \mu}$ the metric equation of motion obtained in \eqref{CovariantPhaseSpace:EquationsOfMotion}
\begin{align}
\nonumber \cE^{\mu \nu} = & - R^{\mu \nu} + \frac{1}{2} g^{\mu \nu} R - 2 \nabla^{\mu} \nabla^{\nu} \Phi + \frac{1}{a_1} \nabla^{\mu} T \nabla^{\nu} T \\
& + \frac{1}{2} g^{\mu \nu} \left( 4 \nabla^2 \Phi - 4 \left( \partial \Phi \right)^2 + 16 - a_1^{-1} \left( \left( \partial T \right)^2 - 4 T^2 + 2 V(T) \right) \right) \, .
\end{align}
Note that in \eqref{EntanglementTargetSpace:HamiltonianSubregion} this is integrated over $\cR$, so that the relevant equation of motion has one component parallel to $\cR$. This is thus a constraint equation in $\cR$. We can now summarize the essential content of \eqref{EntanglementTargetSpace:HamiltonianSubregion} as follows: the Hamiltonian that generates the diffeomorphism along $\zeta$ in $\cR$ is an integral of a constraint equation over $\cR$ plus a boundary term proportional to $e^{-2 \Phi}$.

This can be used in two ways, both of which signal the presence of an area-like term given by that dilaton exponential. The first one is by considering variations between nearby solutions. In that case, we get a classical first law for the local subregion,
\begin{equation}
\delta H_{\zeta}^{\cR} = - \frac{1}{2 \pi} \delta S \, ,
\end{equation}
where
\begin{equation}
\label{EntanglementTargetSpace:AreaTerm}
S = S_L + S_R = 2 \pi a_1 e^{-2 \Phi (\phi_L)} + 2 \pi a_1 e^{-2 \Phi(\phi_R)}
\end{equation}
and we are using that $F_{\zeta} = 0$ on-shell. Similar results have been obtained and discussed extensively in higher-dimensions \cite{Jacobson:2015hqa,Jacobson:2018ahi}. This equation has been suggested to have a thermodynamic interpretation \cite{Jacobson:2018ahi}, in which case we need to attribute the contribution from the boundary points of the interval to the entropy term. Furthermore, adding quantum corrections to the left-hand side in a semiclassical approximation, it is possible to show that a contribution from the modular Hamiltonian of matter fields transforms the right-hand side into a generalized entropy \cite{Jacobson:2018ahi}. We conclude that, if $S$ can be really interpreted as a microscopic entropy (we will comment more on this at the end of the section), such an entropy includes a contribution exponential in the dilaton.

Another hint that we should include such an area term in the generalized entropy arises if we connect our discussion with recent developments discussing operator algebras for generic subregions in gravitational theories \cite{Jensen:2023yxy} -- see \cite{Chandrasekaran:2022cip,Chandrasekaran:2022eqq} for relevant previous works. In this setup, we stay in a region of weak gravity (in our case, far from the tachyon wall) and work perturbatively in the gravitational coupling. When quantizing the resulting theory, we must demand physical operators to be diffeomorphism invariant, which translates to having vanishing commutators with the gravitational constraints. In order for such a condition not to completely trivialize the resulting local algebras, it has become a common practice to introduce an auxiliary observer degree of freedom within the region $\cR$ \cite{Chandrasekaran:2022cip}, universally coupling to the metric through its stress-energy tensor. If we now consider the constraint associated with a diffeomorphism generated by a vector field matching \eqref{EntanglementTargetSpace:CausalDiamondCKV} in a neighborhood of $\cR$, from \eqref{EntanglementTargetSpace:HamiltonianSubregion} we get
\begin{equation}
\label{EntanglementTargetSpace:QuasilocalConstraint}
\int_{\cR} F_{\zeta} = H_{\zeta}^{\cR} + H^{\rm obs} + a_1 \left( e^{-2 \Phi(\phi_L)} + e^{-2 \Phi(\phi_R)} \right) \, .
\end{equation}
with $F_{\zeta}$ giving the quasilocal constraint. Note that the stress tensor within the equation of motion that appears in $F_{\zeta}$ now contains a contribution from the auxiliary observer, which gives rise to the extra term on the right-hand side.

Under some technical assumptions (we refer to the original work \cite{Jensen:2023yxy} for details about them and why they are expected to be true), demanding the above constraint to commute with all operators in subregion $\cR$ gives rise to a type II algebra which has a well-defined notion of entropy (defined up to an additive, state-independent constant). For a general state, this entropy has the form
\begin{equation}
\label{EntanglementTargetSpace:AlgebraicGeneralizedEntropy}
S(\rho_{\hat{\Phi}}) = \langle S_L + S_R\rangle_{\hat{\Phi}} + S_{\Phi}^{\rm mat} + S^{\rm obs} + {\rm const} \, ,
\end{equation}
where $\hat{\Phi}$ is a state of the full system (including the observer), $\Phi$ the corresponding state of the original quantum fields (without the observer), $S^{\rm mat}_{\Phi}$ its entanglement entropy, and $S^{\rm obs}$ the entropy of the observer. This has the form of a generalized entropy, with $S_L + S_R$ as in \eqref{EntanglementTargetSpace:AreaTerm} playing the role of an area-like term.\footnote{It is interesting to note that, due to our previous analysis of the Susskind-Uglum argument, this generalized entropy appears to have UV divergences. This may seem puzzling given that it is obtained from a rigorous algebraic construction. The solution to this puzzle is in the undetermined constant: in two dimensions, the divergence of the matter entropy is just a constant that can be absorbed there. Alternatively, the topological term can be seen as included in that undetermined constant.} The presence of this term in the entropy is a direct consequence of it being present in the quasilocal constraint \eqref{EntanglementTargetSpace:QuasilocalConstraint} \cite{Jensen:2023yxy}.

\subsection{Comments on the different arguments}
\label{subsec:CommentsArguments}

In the previous sections, we have discussed different arguments that hint towards the existence of an area-like term in the entropy of a subregion in the linear dilaton background of the theory \eqref{Review:TargetSpaceAction}. Here we will briefly assess what the arguments really allow to conclude.

The Susskind-Uglum argument has a different status in two-dimensional dilaton gravity theories than it has in higher-dimensional gravity. This is because it renormalizes the coupling of the topological term in the action, as opposed to the coupling of the actual dynamical term $e^{-2\Phi} R$. Thus, we cannot conclude from this construction that an area-like term scaling with the exponential of the dilaton is needed in order to get a generalized entropy with good UV properties.

Still, the argument certainly does not rule out the inclusion of such a term. Furthermore, considering both terms (dilaton dependent and topological) has been essential in recent discussions about islands and how they solve the information problem in two-dimensional setups \cite{Almheiri:2019psf}. In those models, the same feature we discussed is observed: the renormalization only affects the topological coupling. Therefore, even if not directly relevant for the UV properties of the generalized entropy, the dilaton dependent term can be expected to be appear.

More support for this conclusion comes from studying gravitational thermodynamics in the target space theory. We saw that the dilaton dependent contribution to the entropy appears in the first law for black hole horizons, and also in the equivalent relation for causal diamonds in a flat spacetime. We cannot immediately conclude from this result that there is some microscopic entropic quantity that reproduces the entropy-like term obtained in the gravitational first law for a causal diamond (this contrasts with the case of a black hole horizon, where we do expect an interpretation in terms of the number of microstates).\footnote{We will elaborate on this in \secref{sec:subsystems-and-subregions}.} But it is certainly a strong hint that, if a microscopic notion of the entropy of a general subregion exists, it includes a contribution from an area-like term weighted by $e^{-2 \Phi}$.

Similar remarks apply to the algebraic argument leading to the generalized entropy \eqref{EntanglementTargetSpace:AlgebraicGeneralizedEntropy}. In that case, we can make a rigorous construction in the target space theory that connects the proposed entropy with an actual algebraic entropy, but the extra ingredient required -- the inclusion of an observer degree of freedom -- somewhat weakens the result. Indeed, it is not obvious how the dual matrix quantum mechanics would describe this observer, and it is certainly not there in the singlet sector vacuum. Nevertheless, it is reassuring to have a target space construction of the entropy of a subregion in which the dilaton dependent term appears.

To summarize, most arguments that in higher dimensions point towards the existence of an area term in the entropy of a finite subregion seem to have a parallel in the target space theory \eqref{Review:TargetSpaceAction}. The corresponding two-dimensional term depends on the value of the dilaton at the boundaries of the subregion. In the linear dilaton background \eqref{Review:MetricDilatonBackground}, this produces a term exponentially depending on the position of the endpoints of the interval. Given that the double-scaling limit of the matrix quantum mechanics \eqref{Review:MQMAction} is expected to provide a microscopic definition of the theory, it is natural to expect such a contribution to be captured by it. In the remaining sections, we will discuss where such a term can (and cannot) be hidden.

\section{No area term in the singlet sector of the matrix model}
\label{sec:entanglement-MM}

This section explains how to compute entanglement entropy from the matrix model description of the $c=1$ 2d string theory.
We will work entirely in the singlet sector of the matrix model.
In contrast to \cite{Das:1995vj,Hartnoll:2015fca}, we will compute the entanglement entropy of an interval in the target space instead of in the space of eigenvalues.

Computing target space entanglement requires two steps.
First, a bosonization prescription is used to rewrite the fermions on the eigenvalue space in terms of a bosonic collective field variable.
This is discussed in \secref{sec:entanglement-bosonization}.
Then, the correlators of the target space tachyon are related to the correlators of the collective field on eigenvalue space by a nonlocal integral transform which encodes gravitational interactions between tachyons scattering in target space \cite{Natsuume:1994sp}.
In \secref{sec:target-space-entanglement}, we explain how to take into account this nonlocal transformation in a numerical computation of the entanglement entropy.
From the numerical data plotted in \secref{sec:numerical-results}, it can be seen that the nonlocal transformation induces corrections to the entanglement entropy computed in \cite{Hartnoll:2015fca}.
However, for small Newton's constant the corrections are small as well.
For the correlators, the corrections scale with positive powers of $G_N$ \cite{Natsuume:1994sp} and for the entanglement entropy we also find only small corrections in our numerical computation.
Therefore, no area-like term, which would scale as $1/G_N$, is present in the entanglement entropy.
We give an intuitive explanation for why this happens in \secref{sec:nonlocal-transformation}.
Finally, in \secref{sec:eigenvalue-space-entanglement}, we check that our computation methods are consistent with those of \cite{Hartnoll:2015fca} when applied to the eigenvalue space entanglement entropy.

\subsection{Entanglement and bosonization}
\label{sec:entanglement-bosonization}
Let us briefly explain some subtleties in the definition of entanglement in eigenvalue space related to the bosonization map before turning to target space entanglement in \secref{sec:target-space-entanglement}.

The entanglement entropy in the space of eigenvalues of the matrix quantum mechanics is given by $S_A = -\Tr(\rho_A\log\rho_A)$, where $\rho_A$ is the reduced density matrix associated to a subregion $A$ in the (double-scaled) eigenvalue space parametrized by $x \in \mathbb{R}$.
An important conceptual point is that due to the nonlocality of the bosonization map relating \eqref{eq:fermionic-Hamiltonian} and \eqref{eq:bosonic-Hamiltonian}, one might expect the reduced density matrix and thus also the entanglement entropy to depend on whether one works in the bosonic collective field description or in the description in terms of spinless fermions.
This can be most easily argued from the algebraic definition of entanglement entropy, which associates the reduced density matrix to the unique element $\rho_A \in M_A$ of the algebra $M_A$ of operators localized within $A$ which reproduces the same expectation values as the total density matrix $\rho$, $\Tr(\rho_A \cO) = \Tr(\rho \cO)~\forall \cO \in M_A$ \cite{Harlow:2016vwg,OhyaPetz1993}.
Since the Jordan-Wigner transform (equation~\eqref{eq:standard-bosonization} in our conventions) relates the fermionic field at $x=x_1$ to an integral over the bosonic field from an arbitrary but fixed starting point $x=x_0$ to $x=x_1$, the algebra $M_A^\text{bosonic}$ of bosonic operators localized within $A$ will in general differ from that of the fermionic operators $M_A^\text{fermionic}$ localized within the same subregion $A$.

However, for a single interval, which we will restrict to, the entanglement entropy is actually the same in both descriptions.
From the algebraic definition of entanglement entropy, this can be seen by moving the arbitrary interval endpoint $x=x_0$ in the Jordan-Wigner transform to lie at the lower endpoint of the interval $A$.
With this gauge choice, the fermionic operators localized within the subregion $A$ only involve bosonic operators also localized within $A$ and therefore the algebras of local operators are the same, $M_A^\text{bosonic} = M_A^\text{fermionic}$.
This immediately implies that the entanglement entropy is also equal.

For a subregion $A$ consisting of multiple intervals, the different notions of locality in the bosonic and fermionic descriptions do in fact lead to different entanglement entropies.\footnote{It is still possible to compute the entanglement entropy for a fermionic notion of locality in bosonic variables and vice-versa by summing over spin structures (insertions of the fermion number operator within a single interval), see e.g.~\cite{Maric:2020ulk} for a recent discussion. One might also wonder whether such a summation over spin structures is needed for other observables. For the standard correlators of boson or fermion fields, this is irrelevant since the total fermion number is fixed by the size of the matrix and thus the insertion of a spin structure (i.e.~a projector onto even or odd fermion number) is trivial. This argument doesn't apply to the entanglement entropy since the fermion number within a subregion is unconstrained even if the total fermion number is fixed.}
We will, however, not be concerned with such situations since the presence of the area term in the target space entanglement entropy can already be diagnosed from the entanglement for a single interval.
Moreover, the notion of locality we are interested in is that of locality in target space, which is naturally defined in a bosonized description.

\subsection{Target space entanglement entropy}
\label{sec:target-space-entanglement}
Our computation of the entanglement entropy of an interval in target space in based on numerical methods for free theories developed in \cite{Peschel2003,Casini:2009sr,Arias:2018tmw}.
These methods are applicable whenever correlators in the state under consideration factorize into Wick contractions.
As we will argue below, this is true for entangling intervals in the weak-coupling region far away from the tachyon wall.
Because the area-like term dominates in this region and includes an exponential dependence on the interval location, it gives a strong signal that can be clearly distinguished even in a numerical computation.
Therefore, it is sufficient to look at intervals far away from the tachyon wall in order to diagnose the existence of an area term.
To zoom into the weak coupling region, we take a WKB limit $\mu \to \infty$ keeping the time-of-flight coordinate $\tau$ fixed.
Equivalently, this is a limit where the string coupling $g_s$ is small.
The correlators of the tachyon field are related to those of the collective field on eigenvalue space by the nonlocal but linear transformation \eqref{eq:nonlocal-transformation}.
This implies that tachyon correlators factorize into Wick contractions if those of the collective field do so as well.
Factorization of collective field correlators into Wick contractions is assured since the cubic interaction terms in the collective field Hamiltonian \eqref{eq:bosonic-Hamiltonian} drop out in the WKB limit $\mu \to \infty$  and thus the Hamiltonian is quadratic in this limit (see appendix~\ref{sec:details-bosonization-nonlocal-transformation} for a detailed verification).

In more detail, the numerical method works as follows.
Let us first explain the action of the nonlocal transformation onto the tachyon before we come to the discretization.
First, note that the left- and right-moving creation/annihilation operators $\bar{\alpha}_\pm(\omega)$ of the matrix model scalar are related in the WKB limit \cite{Natsuume:1994sp},
\begin{equation}
    \bar{\alpha}_-(\omega) = -\left(\frac{2}{\mu}\right)^{-i \omega}\bar{\alpha}_+(\omega).
    \label{eq:left-right-interaction-scalar-field-modes}
\end{equation}
This interaction between left- and right-moving modes is present even in the limit $\mu \to \infty$ where the cubic interaction terms in the collective field Hamiltonian are negligible and distinguishes the theory from a simple free relativistic scalar field.
Together with the nonlocal transformation, which acts on the creation/annihilation operators as
\begin{equation}
  \alpha_\pm(\omega) = \left(\frac{\pi}{2}\right)^{\pm i \omega/4} \frac{\Gamma(\pm i \omega)}{\Gamma(\mp i \omega)} \bar{\alpha}_\pm(\omega),
\end{equation}
eq.~\eqref{eq:left-right-interaction-scalar-field-modes} implies for the creation/annihilation operators of the target space scalar,
\begin{equation}
    \alpha_-(\omega) = - \left(\frac{2\pi}{\mu^2}\right)^{-i\omega/2}\left(\frac{\Gamma(-i \omega)}{\Gamma(i \omega)}\right)^2 \alpha_+(\omega).
\end{equation}
For the left- and right-moving parts of the target space scalar itself, we get
\begin{equation}
    S(t,\phi)=S_+(t-\phi)+S_-(t+\phi), \quad \text{with} \quad S_-(\phi) = -\int \rd\phi'\, \tilde K(\phi-\phi')S_+(\phi')
    \label{eq:expansion-tachyon-left-right-movers}
\end{equation}
where
\begin{equation}
    \tilde K(\phi) = \int \frac{\rd\omega}{2\pi} e^{i\omega\phi} \left(\frac{\pi}{2}\right)^{-i\omega/2}\left(\frac{\Gamma(-i\omega)}{\Gamma(i\omega)}\right)^2
\end{equation}
is an integration kernel closely related to, but distinct from, the nonlocal transformation \eqref{eq:integration-kernel-nonlocal-transformation}.
It arises from describing the scattering of an incoming tachyon pulse to an outgoing one via transforming to matrix model variables, time evolving in the matrix model and then transforming back to target space variables \cite{Natsuume:1994sp}.
Therefore, in momentum space the square of the leg-pole factor appears.

The discretization procedure works as follows.
We split the target space direction into an equally spaced grid with lattice spacing $a$.
Consider an ordinary free scalar field $\tilde S$ on such a lattice (not to be confused with the matrix model scalar $\bar S$).
The Hamiltonian for this field is discretized as follows,
\begin{equation}
    H = \frac{1}{2} \int \rd\phi \left(\tilde \Pi^2 + (\partial_\phi \tilde S)^2\right) \to \frac{1}{2}\sum_i \left(\tilde \Pi_i^2 + \frac{1}{a^2}(\tilde S_{i+1}-\tilde S_i)^2\right) = \frac{1}{2}\sum_i \tilde \Pi_i^2 + \frac{1}{2}\sum_{i,j} \tilde S_i Q_{ij} \tilde S_j
\end{equation}
where we have introduced a shorthand notation $Q_{ij}$ for the quadratic $\tilde S$ terms.
The correlators for the ordinary free scalar are given by \cite{Casini:2009sr}
\begin{equation}
    \ev{\tilde \Pi_i \tilde \Pi_j} = \frac{1}{2} \left(Q^{1/2}\right)_{ij}, \quad \ev{\tilde S_i \tilde S_j} = \frac{1}{2}\left(Q^{-1/2}\right)_{ij}, \quad \ev{\tilde \Pi_i \tilde S_j} = \frac{i}{2}\delta_{ij}.
    \label{eq:correlators-discretized-ordinary-scalar}
\end{equation}
Given that the only difference between the ordinary free scalar $\tilde S$ and the tachyon $S$ in the WKB limit is the relation between left and right-movers \eqref{eq:expansion-tachyon-left-right-movers}, the field operators for the tachyon in the discretized approach are defined by
\begin{equation}
    S_i = \frac{1}{\sqrt{2}}\left(\tilde S_i - \tilde K_{ij}\tilde S_{-j}\right), \quad \Pi_i = \frac{1}{\sqrt{2}}\left(\tilde \Pi_i - \tilde K_{ij}\tilde \Pi_{-j}\right),
    \label{eq:discretized-target-space-field-operators}
\end{equation}
where
\begin{equation}
    \tilde K_{ij} = a \tilde K(a(i-j)).
    \label{eq:discretized-nonlocal-transformation}
\end{equation}
Note that the minus sign in the subscripts $\tilde S_{-j},\tilde \Pi_{-j}$ in \eqref{eq:discretized-target-space-field-operators} is due to $S(t,\phi)=S_+(t-\phi)+S_-(t+\phi)$, i.e., the sign of the space direction in the argument of the left-/right-movers is reversed.
The correlators of these field operators $S_i,\Pi_i$ can be computed directly from \eqref{eq:correlators-discretized-ordinary-scalar} and \eqref{eq:discretized-target-space-field-operators}.

Due to the fact that higher-point correlators are obtained from the two-point correlators by Wick contractions, we can use methods for computing entanglement entropy in free theories derived in \cite{Peschel2003,Casini:2009sr,Arias:2018tmw}.
In particular, we use a basis of operators which are not canonically conjugate, $[S_i,\Pi_j] \neq i \delta_{ij}$, as can be seen from \eqref{eq:discretized-target-space-field-operators} and $[\tilde S_i, \tilde \Pi_j] = i \delta_{ij}$.
In this setting, the entanglement entropy is given by \cite{Arias:2018tmw}
\begin{equation}
    S = \Tr(V\log|V|)\, ,
    \label{eq:entanglement-entropy-lattice}
\end{equation}
where the $S_i,\Pi_i$ operators localized in the entangling interval have been concatenated into a single vector $f_a$ and the matrix $V$ is defined by
\begin{equation}
    V = -i C^{-1}F\, , \quad \text{where} \quad F_{ab} = \ev{f_a f_b} \quad \text{and} \quad [f_a,f_b]=iC_{ab}.
\end{equation}
This is the result for the entanglement entropy of a target space interval to the far left of the tachyon wall.

\subsection{Numerical results}
\label{sec:numerical-results}
The results of the numerics for the target space entanglement entropy are as follows.
\begin{figure}
    \centering
    \includegraphics[width=0.5\linewidth]{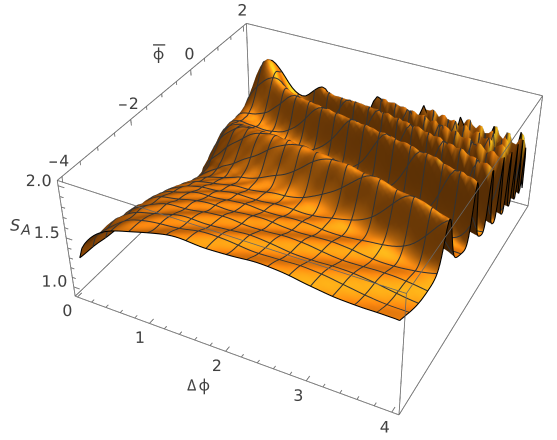}
    \caption{Entanglement entropy for an interval $[\bar\phi-\Delta\phi,\bar\phi+\Delta\phi]$ in target space. Far away from the tachyon wall at $\phi=0$ the entanglement entropy scales logarithmically with both $\Delta\phi$ and $\bar\phi$. For intervals where the endpoint $\bar\phi+\Delta\phi$ comes close to the tachyon wall, the entanglement entropy starts oscillating. The computation used a lattice of 801 points distributed symmetrically around $\phi=0$ with lattice spacing $a=0.025$.}
    \label{fig:3d-plot-numerics-EE}
\end{figure}
The entanglement entropy diverges logarithmically with the lattice spacing $a$, $S_A = -\frac{1}{3}\log(a) + \text{regular part}$.
This divergence is expected for a free (relativistic) scalar field in 1+1 dimensions.
Since we are working in a limit where the cubic interaction terms in the bosonic picture are neglected, finiteness of the entanglement entropy is not visible in our calculation although it would emerge if we were to take into account the full theory.\footnote{To be precise, we expect this to be an order of limits issue. Bosonization is a complete equivalence between fermionic and bosonic descriptions of the same Hilbert space. The ground state in the bosonic description is mapped to the state filled up to the Fermi level in the fermionic description. Therefore, the entanglement entropy for this state has to agree in both theories for any finite value of $\mu$ in the limit of the cutoff $a$ going to zero. We expect a finite answer for the entanglement entropy when first taking the $a \to 0$ and then the $\mu \to \infty$ limit. We are using the reverse order of limits and therefore our result includes a UV divergence. The presence of the area term should not be affected by this issue.}
But since the UV divergence is logarithmic it just gives an additional constant term.
What is of interest to us, namely the dependence of the entanglement entropy on the interval size and location, is meaningful despite the UV divergence.

\begin{figure}
    \centering
    \includegraphics[width=0.48\linewidth]{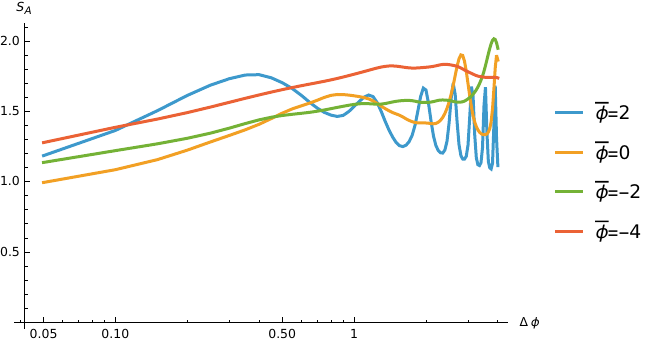}
    \includegraphics[width=0.48\linewidth]{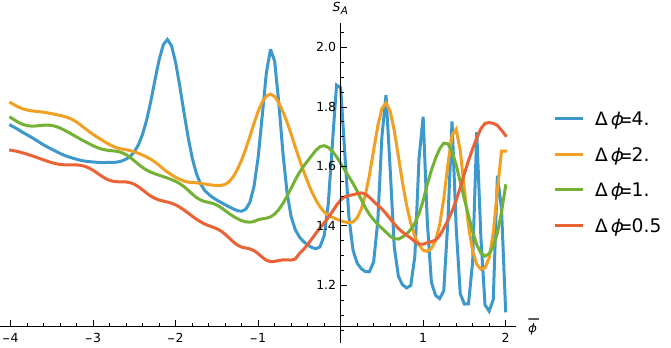}
    \caption{Entanglement entropy at fixed $\bar\phi$ and varying $\Delta\phi$ (LHS) and at fixed $\Delta\phi$ and varying $\bar\phi$ (RHS). The logarithmic scaling of the entanglement entropy for small intervals far away from the tachyon wall is clearly visible on the left hand side (note the logarithmic scale of the $\Delta\phi$ axis). The computation used a lattice of 801 points distributed symmetrically around $\phi=0$ with lattice spacing $a=0.025$.}
    \label{fig:plots-numerics-EE}
\end{figure}

The dependence on the interval size $\Delta\phi$ and interval midpoint $\bar\phi$ is as follows (see Figures~\ref{fig:3d-plot-numerics-EE} and \ref{fig:plots-numerics-EE}).
For small intervals far away from the tachyon wall, i.e., at large negative values of $\phi$, the entanglement entropy scales logarithmically, $S_A \sim c_1 \log(\Delta\phi) + c_2 \log(|\bar\phi|)$.
A similar kind of logarithmic scaling was found for the eigenvalue space entanglement in \cite{Hartnoll:2015fca}; however, for the target space entanglement the coefficients $c_1,c_2$ are both positive while \cite{Hartnoll:2015fca} found $c_1 = - c_2 = 1/3$.\footnote{The exact coefficients are difficult to determine from the available numerical data. The best fit and consistency with the cutoff scaling $\sim -1/3 \log(a)$ suggests $c_1 = 1/3-c_2 \approx 1/6$.}
For intervals where one of the endpoints comes close to the tachyon wall, we start to see oscillations w.r.t.~$\Delta\phi$ and $\bar\phi$ which become increasingly fast the more the interval reaches into the strongly coupled region.
However, at this point the numerical result is no longer trustworthy since neglecting the interactions in the bosonic description is only valid far away from the tachyon wall.

We find no trace of an area-like term which would scale exponentially with $\Delta\phi$ and $\bar\phi$.
In particular, the area-like term should dominate over the matter contribution to the entanglement entropy far away from the tachyon wall.
However, our numerical results agree better and better with the matter contribution the further away from the tachyon wall we are.
Therefore, we conclude that the entanglement entropy for the $c=1$ matrix model does not contain an area-like term if we restrict only to the singlet sector.
Including the effect of the nonlocal interaction which was neglected in \cite{Hartnoll:2015fca} produces subleading effects in $G_N$ but not a leading order $1/G_N$ area-like term.

\subsection{A closer look at the nonlocal transformation}
\label{sec:nonlocal-transformation}
\begin{figure}
    \centering
    \includegraphics[width=0.75\linewidth]{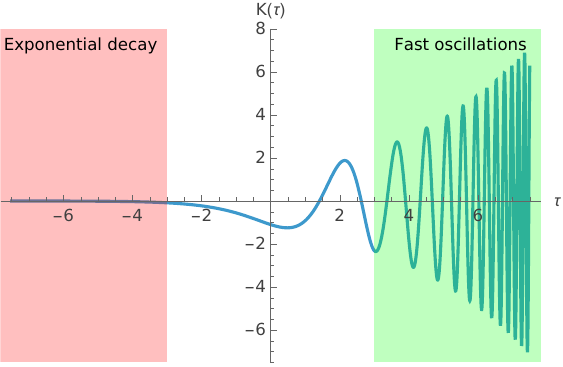}
    \caption{Plot of the nonlocal transformation $K(\tau)$. The function falls off exponentially to the left in the red-shaded region while it oscillates with increasing frequency to the right in the green-shaded region. Only the region close to $\tau=0$ contributes significantly as other contributions are either exponentially small or average out. Thus, the transformation $K(\tau)$ only introduces a small amount of nonlocality with the largest contributions coming from close to the point where the eigenvalue and target space coordinates are equal.}
    \label{fig:nonlocal-transformation}
\end{figure}
To get more intuition as to why the corrections to the entanglement entropy from the nonlocal transformation are small, it is instructive to plot the function $K(\tau)$ relating the target space and matrix model scalar fields $S$ and $\bar S$ (see Figure~\ref{fig:nonlocal-transformation}).
The asymptotic behavior of this functions is given by \cite{Natsuume:1994sp}
\begin{equation}
    \begin{aligned}
     K(\tau) &\sim - \left(\frac{\pi}{2}\right)^{-1/4}e^\tau, \quad \tau \to -\infty\\
             &\sim \left(\frac{\pi}{2}\right)^{-1/16} \frac{e^{\tau/4}}{\sqrt\pi} \cos\left(2(2/\pi)^{1/8}e^{\tau/2}+\pi/4\right), \quad \tau \to +\infty.
    \end{aligned}
\end{equation}
Contributions to the integral transform $S(\phi) = \int \rd\tau K(\phi-\tau)\bar S(\tau)$ from $\phi - \tau \to -\infty$ are exponentially suppressed while contributions from $\phi - \tau \to +\infty$ largely average out in the integral since $K(\tau)$ oscillates with exponentially increasing frequency as $\tau$ becomes large.
Therefore, the main contribution to the nonlocal transformation comes from the region $\phi \approx \tau$.
In other words, the amount of nonlocality introduced by the integral transform is small.
In order to find an area-like term in the entanglement entropy, a drastic change would have needed to occur after applying the nonlocal transformation with the area-like term becoming increasingly dominant as the interval moves to the left of the tachyon wall into the small coupling region.
This is clearly not implemented by the transformation introduced from $K(\tau)$, which explains intuitively why no area-like term emerges from the nonlocal transformation.

\subsection{Eigenvalue space entanglement}
\label{sec:eigenvalue-space-entanglement}
A useful cross check of our method is to compare the entanglement entropy for an interval in eigenvalue space with the corresponding results from \cite{Hartnoll:2015fca} obtained in the fermionic description of the singlet sector.
In our bosonic description, the eigenvalue space entanglement is computed using the same formulas \eqref{eq:entanglement-entropy-lattice} and \eqref{eq:discretized-target-space-field-operators} as for the target space, except that now the nonlocal transformation trivializes, meaning that instead of \eqref{eq:discretized-nonlocal-transformation} we have
\begin{equation}
    \tilde K_{ij} = \delta_{ij}.
\end{equation}
To the extent that is possible to compare, our results agree with those of \cite{Hartnoll:2015fca}.

A one-to-one comparison is not possible since both computations are approximate.
In \cite{Hartnoll:2015fca}, the entanglement entropy is expressed in terms of an infinite sum over cumulants of the particle number distribution which is subsequently approximated by the first term.
Moreover, a WKB limit is employed in order to compute the first cumulant.
In our method, the interaction terms in the collective field Hamiltonian \eqref{eq:bosonic-Hamiltonian} are neglected and the entanglement entropy is computed by a numerical approximation.
The latter approximation can be made as precise as desired by decreasing the lattice spacing $a$ and thus does not impede a comparison to the result from the fermionic description.
Neglecting the interaction terms is justified in the WKB limit, which was also employed in \cite{Hartnoll:2015fca}.

One feature of the result that can be compared between our approach and that of \cite{Hartnoll:2015fca} is the logarithmic scaling for small interval size $\tau_2-\tau_1 \to 0$ which is argued in \cite{Hartnoll:2015fca} to be robust against including higher cumulant contributions.
We indeed find a logarithmic scaling $S \sim c_1 \log(\tau_2-\tau_1)$ in the eigenvalue entanglement entropy.\footnote{The scaling coefficient $c_1$ is difficult to extract numerically.
The best fit suggests $c_1 \approx 1/4$ which disagrees with $c_1 = 1/3$ from \cite{Hartnoll:2015fca}. This might be due to the different approximation schemes employed in both computations. The only feature that can be reliably extracted is the presence of the logarithmic term and not the precise coefficient.}
Moreover, both the result of \cite{Hartnoll:2015fca} as well as our numerical computation show a $S \sim c_2 \log(\tau_1 \tau_2)$ scaling at $\tau_{1,2} \to 0$ where the interval endpoint in eigenvalue space almost approaches the top of the inverted harmonic oscillator potential.\footnote{A fit of our numerical result gives $c_2 \approx 1/6$ in agreement with \cite{Hartnoll:2015fca}, although higher numerical precision would be needed to precisely confirm or exclude an agreement and this scaling might be modified by higher cumulant contributions to the result of \cite{Hartnoll:2015fca}.}

In summary, these results indicate that our computation method for the eigenvalue space entanglement entropy is compatible with that of \cite{Hartnoll:2015fca}.

\section{Comments on non-singlets and generic subregions}
\label{sec:final-comments}

Having shown that the nonlocal leg-pole transformation does not generate an area-like term in the entanglement entropy of a target space subregion, the only remaining place to find it seems to be some contribution from the non-singlet sector of the matrix quantum mechanics. If this is the case, it would resonate well with previous results that show that a description of the quintessential gravitational feature of the theory -- the black hole solutions \cite{Mandal:1991tz,Witten:1991yr,Dijkgraaf:1991ba} -- is possible in the matrix model only if we include non-singlets \cite{Kazakov:2000pm,Karczmarek:2004bw}. In this section, we provide some preliminary and speculative comments about the role of non-singlets for the computation of the entropy of a subregion, leaving a more detailed analysis for future work. We will end with some comments that must be kept in mind in case even non-singlets cannot explain the gravitational area law for the entropy of a target space subregion.

\subsection{Non-singlets}
\label{sec:non-singlets}

The target space linear dilaton background (with the tachyon wall), possibly with some low-energy excitations, is well described in the matrix model using singlet sector states. One may naively conclude from this that the non-singlet sectors of the theory do not play any role in the computation of the entanglement entropy of a target space subregion: if the state has no support on such non-singlet states, they should not contribute to the density matrix obtained after tracing out part of the degrees of freedom.

This is not entirely correct, because we can involve the non-singlet degrees of freedom in our definition of the subregion. An alternative way to say this is that in order to define a subsystem of the matrix quantum mechanics we must pick a subalgebra of operators. This is what was done (perhaps implicitly) in previous works \cite{Das:1995vj,Hartnoll:2015fca}, and what underlies the calculation in section \ref{sec:entanglement-MM}: we choose the subalgebra of operators localized to an interval of the eigenvalue space and we compute the entanglement entropy of such subalgebra in the given singlet state. This subalgebra does not mix the singlet sector with the non-singlet sectors (it only acts on the eigenvalues), so indeed non-singlets are totally irrelevant.

In principle, it is possible that defining an algebra that involves the non-singlet degrees of freedom we get a different result for the entropy, even if the state under consideration is a singlet. We present a toy model of this phenomenon in appendix \ref{app:EntanglementNonInvariant}, where for simplicity we look at a system of $N$ distinguishable particles. The Hilbert space of this system contains an invariant (singlet) sector under permutation of the particles: this is the subspace of symmetric wavefunctions, relevant for the description of identical bosons. There are two different notions of a subalgebra associated to a subregion $R$ that we can define: in one, relevant if we care only about the singlet sector, we consider symmetrized combinations of operators in $R$; in the other, relevant in the full theory of distinguishable particles, we allow any operator in $R$. Even if we consider a fixed singlet state, the result for the entropy is shown to depend on the algebra chosen, with a relatively simple relation between both entropies determined by the symmetry of the state \eqref{EntanglementNonInvariant:RelationEntropies}:
\begin{equation}
\label{non-singlets:EntropiesRelation}
S(R) = S_{\rm symm.}(R) + \sum_{k=0}^N p_{0,k} \log \binom{N}{k} \, .
\end{equation}
Here, the left hand side is the entropy for the algebra of all operators in $R$, while the right hand side is the entropy for the algebra of symmetrized operators in $R$ plus an extra sum, in which $p_{0,k}$ is the probability to find $k$ particles in region $R$ in the given symmetric state.

If we translate this to the $c=1$ matrix quantum mechanics, we have always been computing the entropy for an algebra of singlet sector operators, a quantity analogous to $S_{\rm symm.}(R)$. We can hope that choosing a more general algebra that involves non-singlets (but which reduces to the usual one when projecting to the singlet sector) there is an extra term that reproduces the gravitational area law.

Fully elucidating whether this is possible is left for future work, but we want to offer here some preliminary comments about the difficulties one encounters in following this direction. The matrix $M$ can be parametrized in terms of its $N$ eigenvalues, collectively denoted by $\Lambda$, and $N^2 - N$ ``angular'' variables that specify the unitary transformation needed to diagonalize $M$ (we denote these by $\Omega$). The $SU(N)$ symmetry acts only by rotation of the angles. A general wavefunction is $\psi(\Lambda, \Omega)$, while a singlet state has no dependence on $\Omega$.\footnote{A concrete example may help here, so let us take $N=2$. A Hermitian matrix can be parametrized as
\begin{equation*}
M = \frac{1}{2} \begin{pmatrix}
    \lambda_1 + \lambda_2 +(\lambda_1-\lambda_2) \cos \theta && (\lambda_1 - \lambda_2) \sin \theta e^{-i \phi} \\
    (\lambda_1 - \lambda_2) \sin \theta e^{i \phi} && \lambda_1 + \lambda_2 - (\lambda_1-\lambda_2) \cos \theta
\end{pmatrix} \, ,
\end{equation*}
where $\lambda_1 \geq \lambda_2$ are the eigenvalues and $\theta \in (0, \pi)$, $\phi \in (0, 2\pi)$ are the angles parametrizing an $\mathbb{S}^2$. An $SU(2)$ transformation (we really have to quotient out the center, so it is in $SO(3) \cong SU(2)/\mathbb{Z}_2$) acts by rotating the angles in the usual way. The wavefunction is a function $\psi(\lambda_1, \lambda_2, \theta, \phi)$, the angles can be expanded in spherical harmonics and the singlet sector corresponds to the $l = 0$ component.
}

To choose the subalgebra that characterizes the subregion in target space, the most natural approach is to keep the same division of the eigenvalue space that we had when ignoring non-singlets. So we select operators that act non-trivially only on a certain interval of the eigenvalue space. But we still have the freedom to choose how the operators act on the angular variables $\Omega$ (with the restriction that the full set of operators must form an algebra). If we pick operators that do not act at all on the $\Omega$, these leave the singlet sector invariant and we are back to the previous results. We can come up with other alternatives, for example we can allow the operators to act pointwise on the angles
\begin{equation}
\label{non-singlets:OperatorExample}
(\hat{O} \psi )(\Lambda,\Omega) = \int_R \rd \Lambda' \, O(\Omega; \Lambda, \Lambda') \psi( \Lambda', \Omega) \, ,
\end{equation}
where the integral over $R$ should be understood as restricting to the family of operators considered in eigenvalue space. This is of course just one among many possibilities.

The issue with simple proposals like \eqref{non-singlets:OperatorExample} is that they do not relate the action on the angles to the region chosen in eigenvalue space. Thus, when computing the entropy for the algebra in a singlet state, we may get extra terms as in \eqref{non-singlets:EntropiesRelation}, but they will not depend on the position of the boundary points of the interval. Given that the area term we are after does depend on the positions of the boundary points (through the dilaton $e^{-2 \Phi(x)}$), we will need to choose our algebra by introducing a non-trivial relation between the action on the angles and the region in eigenvalue space. We leave for future work to clarify whether such a construction can give a contribution to the entropy that matches the target space area term.

\subsection{What if non-singlets do not produce an area term?}
\label{sec:subsystems-and-subregions}

It is in principle possible that, even considering more general algebras, we are not able to find a contribution to the entropy in the matrix quantum mechanics that matches the area term of the target space theory. If this is the case, what should we conclude? Does it mean that the matrix quantum mechanics is not able to capture the entropy of subregions of the target space, thus casting doubts on this aspect of the $c = 1$ duality?

We want to end here with a short comment about another possibility, which rests on an implicit assumption made in both the present and previous works on the topic of entanglement in the $c=1$ model \cite{Das:1995vj,Hartnoll:2015fca}. We have always taken for granted that \textit{any} subregion (or at least any interval) of target space should correspond to a well-defined subsystem in the microscopic quantum-mechanical description. While this is in the spirit of recent works that introduced operator algebras for generic subregions in the presence of observers \cite{Jensen:2023yxy}, it is far from guaranteed. In fact, other authors have recently introduced the notion of the generalized entanglement wedge $W_R$ of an arbitrary region $R$ \cite{Bousso:2022hlz,Bousso:2023sya}: $W_R$ is the region containing $R$ with smallest generalized entropy.\footnote{There is a technical topological assumption: both $R$ and $W_R$ are demanded to be open and equal to the interior of their respective closures.} Much as in AdS/CFT only entanglement wedges correspond to boundary (spatial) subsystems, the proposal is that well-defined microscopic subsystems are only associated to generalized entanglement wedges. This is also supported by recent results that show how certain properties of generalized entanglement wedges arise naturally if their generalized entropies correspond to entropies of subalgebras of operators in a quantum system \cite{Sahu:2025upe}.

If this proposal holds, we should restrict our analysis to generalized entanglement wedges in the target space of the linear dilaton background. What are these? Consider a target space interval $(\phi_L, \phi_R)$ and look for a region containing it that minimizes the generalized entropy. The exponential dependence of the area term will push the right boundary towards larger values, where it becomes smaller. This tends to take the generalized entanglement wedge to the strong coupling region. While a more careful analysis is needed, and one should always keep in mind that the validity of the generalized entanglement wedge proposal is not proven in this setup, we wanted to highlight here the possibility that not every target space region is associated with a valid microscopic subsystem.

\acknowledgments

We would like to thank Sumit Das and Alex Frenkel for many enlightening discussions about entanglement in the $c=1$ model. We also thank Sean Hartnoll, Jeremy van der Heijden and Mark Van Raamsdonk for helpful discussions; and Vijay Balasubramanian for collaboration on related topics. Work at VUB is supported by FWO-Vlaanderen projects G012222N and G0A2226N, and by the VUB Research Council through the Strategic Research Program High-Energy Physics. MG is supported by FWO-Vlaanderen through a Junior Postdoctoral Fellowship 1238224N. MP is supported by FWO-Vlaanderen through a Doctoral Fellowship 1178725N. AVL is supported by the National Science and Engineering Research Council of Canada (NSERC) and the Simons Foundation via a Simons Investigator Award.

\appendix

\section{Covariant phase space analysis of the target space theory}
\label{app:CovariantPhaseSpace}

Covariant phase space methods in gravity have a long history \cite{Crnkovic:1986ex,Lee:1990nz,Iyer:1994ys,Barnich:2001jy}, see \cite{Harlow:2019yfa,Chandrasekaran:2021vyu} for recent reviews. In this appendix, we obtain some results valid for the two-dimensional target space effective theory \eqref{Review:TargetSpaceAction}. We will follow the notation and conventions of \cite{Harlow:2019yfa}, and introduce only the general results we need, applying them to our theory. In short, our goal is to find a Hamiltonian function that generates diffeomorphism transformations in phase space. Conveniently applied, this provides access to first law identities in several situations, most notably in the presence of black hole horizons.

The construction starts from a careful analysis of the variational problem defined by the action. Write the Lagrangian form of the theory as
\begin{equation}
\label{CovariantPhaseSpace:Lagrangian}
L = \cL \, \epsilon \, , \qquad \cL \equiv \frac{1}{2} e^{-2 \Phi} \left[ a_1 \left( R + \frac{16}{\alpha'} + 4 \left( \partial \Phi \right)^2 \right) - \left( \partial T \right)^2 + \frac{4}{\alpha'} T^2 - 2 V(T) \right] \, ,
\end{equation}
where $\epsilon$ is the volume form. A variation of the Lagrangian produces
\begin{equation}
\delta L = \frac{1}{2} e^{-2 \Phi} \left[ a_1 \cE^{\mu \nu} \delta g_{\mu \nu} + 2 a_1 \cE^{\Phi} \delta \Phi + 2 \cE^T \delta T \right] \epsilon + \rd \Theta \, ,
\end{equation}
with the following equations of motion
\begin{subequations}
\label{CovariantPhaseSpace:EquationsOfMotion}
\begin{align}
\cE^{\mu \nu} + \frac{1}{2} g^{\mu \nu} \cE^{\Phi} & =  - R^{\mu \nu} - 2 \nabla^{\mu} \nabla^{\nu} \Phi + \frac{1}{a_1} \nabla^{\mu} T \nabla^{\nu} T \, , \\
\cE^{\Phi} & = - R - 4 \nabla^2 \Phi + 4 \left( \partial \Phi \right)^2 - \frac{16}{\alpha'} + \frac{1}{a_1} \left( \left( \partial T \right)^2 - \frac{4}{\alpha'} T^2 + 2 V(T) \right) \, , \\
\cE^T & = \nabla^2 T - 2 \nabla_{\mu} \Phi \nabla^{\mu} T + \frac{4}{\alpha'} T - V'(T) \, .
\end{align}
\end{subequations}
The relevant piece for the analysis that follows is actually the boundary term
\begin{equation}
\Theta = \theta \cdot \epsilon \, ,
\end{equation}
where
\begin{equation}
\label{CovariantPhaseSpace:Theta}
\theta^{\alpha} = \frac{a_1 e^{-2 \Phi}}{2} \left[ 8 \delta \Phi \nabla^{\alpha} \Phi + 2 g^{\mu [\alpha} \nabla^{\nu]} \delta g_{\mu \nu} + 4 \left( g^{\mu [\alpha} \nabla^{\nu]} \Phi \right) \delta g_{\mu \nu} - \frac{2}{a_1} \delta T \nabla^{\alpha} T \right] \, .
\end{equation}

Given that we started from a covariant theory, one immediate consequence of the covariant phase space analysis is that the current $j_{\xi}^{\mu}$ given by
\begin{equation}
j_{\xi}^{\mu} = \theta^{\mu}[\delta_{\xi} \psi] - \xi^{\mu} \cL \, 
\end{equation}
is conserved on-shell, $\nabla_{\mu} j_{\xi}^{\mu} = 0$. Here, $\xi$ is a vector field that we view as a diffeomorphism generator, $\delta_{\xi} \psi = \cL_{\xi} \psi$ means all variations are converted to diffeomorphism variations generated by the Lie derivative $\cL_{\xi}$, and $\cL$ is the Lagrangian scalar in \eqref{CovariantPhaseSpace:Lagrangian}. In our theory, we obtain
\begin{equation}
j_{\xi}^{\mu} = - a_1 e^{-2\Phi} \xi_{\lambda} \cE^{\lambda \mu} + \nabla_{\lambda} \left[ \frac{a_1 e^{-2 \Phi}}{2} \left( 2 \nabla^{[\lambda} \xi^{\mu]} - 8 \xi^{[\lambda} \nabla^{\mu]} \Phi \right) \right] \, .
\end{equation}
It is convenient to view this as a $1$-form, $J_{\xi} = j_{\xi} \cdot \epsilon$, with $\epsilon$ being the volume two-form. The first piece is proportional to the metric equation of motion, while the fact that we are working with a fully covariant theory makes the second piece a total derivative. We write this as
\begin{equation}
J_{\xi} = F_{\xi} + \rd Q_{\xi} \, , 
\end{equation}
with
\begin{equation}
\label{CovariantPhaseSpace:ConstraintGenerator}
F_{\xi} = f_{\xi} \cdot \epsilon \, , \qquad f_{\xi}^{\mu} = - a_1 e^{-2\Phi} \xi_{\lambda} \cE^{\lambda \mu} \, ,
\end{equation}
proportional to the metric equation of motion, and
\begin{equation}
\label{CovariantPhaseSpace:NoetherCharge}
Q_{\xi} = - \frac{e^{-2 \Phi}}{2} a_1 \left( \star \rd \xi - 4 \star \left( \xi \wedge \rd \Phi \right) \right) = - \frac{e^{-2 \Phi}}{2} a_1 \epsilon^{\mu \nu} \left(\nabla_{\mu} \xi_{\nu} - 4 \xi_{\mu} \nabla_{\nu} \Phi \right) \, .
\end{equation}

Consider now a spatial region $\cR$, which can be viewed as a Cauchy slice for its domain of dependence. Starting from $\Theta$, which is a one-form in phase space (i.e., linear in field variations), we can define a symplectic form in phase space as
\begin{equation}
\label{CovariantPhasSpace:SymplecticForm}
\Omega = \int_{\cR} \omega \, , \qquad \omega = \delta \Theta \, ,
\end{equation}
where $\delta$ should be understood as exterior differentiation in phase space.\footnote{There are boundary ambiguities in this construction and several proposals exist to deal with them \cite{Harlow:2019yfa,Chandrasekaran:2021vyu}. We will ignore this issue here, since our final result will not depend on them.} It is often useful to think of $\Omega$ (or $\omega$) as evaluated on two particular field variations, writing then $\Omega[\delta \psi, \delta' \psi]$, collectively denoting by $\psi$ the dynamical fields. This is technically a contraction of the two-form on phase space $\Omega$ with two tangent vectors, which correspond to the field variations $\delta \psi$ and $\delta' \psi$. Given a vector field $\xi$, consider the quantity
\begin{equation}
\label{CovariantPhaseSpace:HamiltonianToSymplecticForm}
\hat{\delta} H_{\xi} \equiv \Omega[\delta \psi, \delta_{\xi} \psi] = \int_{\cR} \delta F_{\xi} + \int_{\partial \cR} \left(\delta Q_{\xi} - \xi \cdot \Theta \right) \, ,
\end{equation}
where the final identity between the symplectic form evaluated for one diffeomorphism variation and the integrals over $\cR$ and $\partial \cR$ follows from the covariant phase space analysis -- note that, since we are in two dimensions, the integral over $\partial \cR$ is just a sum over boundary points with appropriate orientations. The hat emphasizes that this is not a priori a total variation (an exact form in phase space). A particularly important case of the previous identity, forming the basis of the gravitational first law, arises when we take $\xi$ to be a Killing field of the background and $\delta$ a variation between solutions. Then $\delta_{\xi} \psi = 0$, so the left-hand side vanishes, and $\delta F_{\xi} = 0$. As a consequence,
\begin{equation}
\label{CovariantPhaseSpace:FirstLaw}
0 = \int_{\partial \cR} \left(\delta Q_{\xi} - \xi \cdot \Theta \right) \, .
\end{equation}

More generally, for a generic vector field $\xi$ (not necessarily Killing), in many important situations $\hat{\delta} H_{\xi}$ in \eqref{CovariantPhaseSpace:HamiltonianToSymplecticForm} turns out to be exact, so that we can view it as the variation of a function $H_{\xi}$. A particularly relevant example for us is when $\xi$ vanishes at $\partial \cR$, in which case we can define
\begin{equation}
\label{CovariantPhaseSpace:HamiltonianIntegrated}
H_{\xi} = \int_{\cR} F_{\xi} + \int_{\partial \cR} Q_{\xi} \, .
\end{equation}
In many other cases, boundary conditions at $\partial \cR$ allow to write $\xi \cdot \Theta$ as a variation, obtaining a corresponding object analogous to $H_{\xi}$. The relation between $H_{\xi}$ and the symplectic form \eqref{CovariantPhaseSpace:HamiltonianToSymplecticForm} ensures that $H_{\xi}$ generates, via Poisson brackets, diffeomorphism transformations in any dynamical field,
\begin{equation}
\delta_{\xi} \psi = \cL_{\xi} \psi = \{ \psi, H_{\xi} \} \, .
\end{equation}
Thus, $H_{\xi}$ is the Hamiltonian associated to the infinitesimal diffeomorphism generated by $\xi$. Note that the first piece in $H_{\xi}$, containing $F_{\xi}$ and integrated over $\cR$, is proportional to the gravitational constraint equations on $\cR$. This is to be expected: in gravity, diffeomorphism transformations are generated by the constraint equations up to boundary terms. When these diffeomorphisms are actual gauge transformations of the theory, such as when $\xi$ is compactly supported and does not reach $\partial \cR$, the constraint equations are the appropriate generators.

If $\cR$ is part of a larger Cauchy slice $\Sigma$, so that $\Sigma = \cR \cup \bar{\cR}$, the above procedure can be repeated for the whole $\Sigma$. In that case, it is often useful to view the Hamiltonian as having two terms, one coming from each piece of $\Sigma$:
\begin{equation}
H_{\xi}^{(\Sigma)} = H_{\xi}^{(\cR)} + H_{\xi}^{(\bar{\cR})} \, ,
\end{equation}
with each term on the right of the form \eqref{CovariantPhaseSpace:HamiltonianIntegrated}, integrating over the appropriate region. Note that we view the common boundary between $\partial \cR$ and $\partial \bar{\cR}$ with opposite orientation (the normal is always assumed outward pointing), so that the corresponding pieces in the boundary terms cancel, leaving only a contribution from $\partial \Sigma$ in the full Hamiltonian.

As a final comment, all the previous discussion can be repeated for a slightly different Lagrangian in which we add a topological term
\begin{equation}
L \rightarrow L + b_B R \, \epsilon \, .
\end{equation}
This does not alter the equations of motion, but it gives a contribution to the boundary term that propagates to $Q_{\xi}$
\begin{equation}
Q_{\xi} \rightarrow Q_{\xi} - b_B \epsilon^{\mu \nu} \nabla_{\mu} \xi_{\nu} \, .
\end{equation}
This adds a constant term to entropy computations.

\section{Cancellation of divergences in the generalized entropy}
\label{app:GeneralizedEntropy}

This appendix contains a detailed calculation that shows the cancellation of divergences in the generalized entropy of the theory \eqref{Review:TargetSpaceAction} (with a topological term added), as discussed in section \ref{subsec:UVFiniteness}. The argument parallels the one developed in \cite{Demers:1995dq} for a four-dimensional theory. With respect to the main text, the main novelty is that we will be careful with the definition of the UV regulator, making sure that we use the same one when computing the renormalization of the gravitational coupling and the von Neumann entropy of the subregion. In the end, in two dimensions this is not very important: the divergences are logarithmic, so they are not affected by rescaling the regulator.

We start by computing the one-loop effective action obtained by integrating out the tachyon. It is enough to work with the quadratic part of its action
\begin{equation}
\cI_T = \frac{1}{2} \int \rd^2 x \sqrt{-G} e^{-2 \Phi} T \left( \nabla^2 - 2 \nabla^{\mu} \Phi \nabla_{\mu} + \frac{4}{\alpha'} \right) T \, ,
\end{equation}
where we have performed an integration by parts to make manifest the second-order differential operator. We compute the UV-divergent part of the effective action $W$ via heat-kernel techniques, see \cite{Vassilevich:2003xt} for a review and all the technical results needed in what follows. Rescaling the tachyon field\footnote{The following manipulations assume that we are integrating out the tachyon field with the measure $\int DT \exp(-\int \rd^2x \sqrt{G} e^{-2\Phi} T^2 )=1$. This is natural in this context: $\tilde{T}$ is the massless scalar that propagates in the linear-dilaton background. In any case, more general normalizations do not affect the renormalization of the gravitational term in the action -- see section 7.1 of \cite{Vassilevich:2003xt}.}
\begin{equation}
\tilde{T} = e^{-\Phi} T \, ,
\end{equation}
the action can be written in terms of a second-order differential operator in a standard form
\begin{equation}
\cI_T = \frac{1}{2} \int \rd^2 x \sqrt{-G} \, \tilde{T} \cD \tilde{T} \, ,
\end{equation}
where
\begin{equation}
\cD = - G^{\mu \nu} \nabla_{\mu} \nabla_{\nu} - E \, , \qquad E \equiv - (\partial \Phi)^2 + \nabla^2 \Phi + \frac{4}{\alpha'} \, .
\end{equation}

The effective action is given by $W = (1/2) \log \det \cD$. This can be written (up to an additive constant) as an integral of the heat-kernel trace
\begin{equation}
W = - \frac{1}{2} \int_0^{\infty} \frac{\rd t}{t} K(t; \cD) \, , \qquad K(t; \cD) = {\rm Tr} \; e^{-t \cD} \, ,
\end{equation}
and this expression is well-suited to isolate UV-divergences, which come from the limit $t \to 0^+$. A series expansion of the heat-kernel trace can be obtained in that limit \cite{Vassilevich:2003xt},
\begin{equation}
K(t; \cD) = \frac{1}{t} \left( a_0(\cD) + t \, a_2(\cD) + \dots \right) \, ,
\end{equation}
where we have only kept terms that lead to UV-divergences, and the coefficients are
\begin{equation}
a_0(\cD) = \frac{1}{4 \pi} \int \rd^2 x \, \sqrt{G} \, , \qquad a_2(\cD) = \frac{1}{4 \pi} \int \rd^2 x \, \sqrt{G} \left( \frac{R}{6} + E \right) \, .
\end{equation}
The first coefficient produces a one-loop renormalization of the cosmological constant. The result \eqref{EntanglementTargetSpace:EffectiveActionCurvature} can be read directly from $a_2(\cD)$, integrating the heat-kernel trace with a lower cutoff in $t$ given by $\epsilon^2$ (note that $t$ has units of length squared).

This is enough to obtain the results presented in section \ref{sec:EntanglementTargetSpace}. Essentially, since the UV divergence multiplying the Ricci scalar is logarithmic, we do not need to be very careful with the precise definition of the regulator: the divergence from the renormalization of the topological coupling cancels the one coming from the tachyon entanglement entropy in any case. Still, it is instructive to implement a consistent regularization method throughout the whole calculation, both for the coupling renormalization and for the matter entanglement entropy. We do this in the remaining part of this appendix.

Let us then use the Pauli-Villars regularization method. In two dimensions, this is done by introducing three extra fields, two anticommuting (so that they enter in $W$ with the opposite  sign) with mass $\Lambda$ and one commuting with mass $\sqrt{2} \Lambda$ \cite{Birrell:1982ix}. By mass here we mean that we modify $E \to E - m_{\Lambda}^2$, with $m_{\Lambda}^2$ the appropriate value for each of the fields. This produces a factor of $e^{- m_{\Lambda}^2 t}$ in the heat-kernel trace. Focusing only on the piece of the effective action multiplying $R$, we get
\begin{equation}
W \supset - \frac{1}{48 \pi} \int \rd^2 x \sqrt{G} R \int_{0}^{\infty} \frac{\rd t}{t} \left( 1 - 2 e^{- \Lambda^2 t} + e^{-2 \Lambda^2 t} \right) \, .
\end{equation}
This integral has an IR divergence that can be regularized by giving an infinitesimal mass to the tachyon. In any case, for our purposes we only care about the UV-divergent piece, which can be unambiguously extracted to be
\begin{equation}
W \supset - \frac{1}{48 \pi} \log(\Lambda^2) \int \rd^2 x \sqrt{G} R \, .
\end{equation}

As discussed in \eqref{EntanglementTargetSpace:TopologicalTerm}, to absorb this divergence we need to add a topological piece to the target space action
\begin{equation}
- b_B\int \rd^2 x \sqrt{G} \, R \, ,
\end{equation}
so that the previous computation gives the renormalized coupling
\begin{equation}
b = b_B + \frac{1}{48 \pi} \log (\Lambda^2) \, .
\end{equation}

We now proceed to compute the generalized entropy of a target space subregion $\cR$ in the linear-dilaton background. In order to be able to compute the matter entropy, we will look in this appendix at a half-space in the weakly coupled region, namely $\phi \in (-\infty, \phi_R)$. The generalized entropy is given by
\begin{equation}
S_{\rm gen} = 4 \pi b_B + 2 \pi a_1 e^{-2\Phi(\phi_R)} + S_{\rm mat} \, .
\end{equation}
Note that there is a single boundary point now. To obtain $S_{\rm mat}$ in the theory with the extra Pauli-Villars fields, we will view the region $\cR$ as a Rindler wedge with metric
\begin{equation}
\rd s^2 = - \rho^2 \rd T^2 + \rd \rho^2 \, , \qquad \rho > 0 \, ,
\end{equation}
so that the fields restricted to $\cR$ are in a thermal state with inverse temperature $\beta = 2 \pi$. We can then identify $S_{\rm mat}$ as the thermal entropy of a Rindler wedge at $\beta = 2\pi$: we proceed now to compute this quantity in the theory with the Pauli-Villars regulators.

Each of the fields satisfies in the linear-dilaton background a massive scalar wave equation. Let us collectively denote them by $\tilde{f}_{i}$ for $i \in \{1,2,3,4\}$, with masses $m_1 = 0$, $m_2 = m_3 = \Lambda$ and $m_4 = \sqrt{2} \Lambda$. We introduce also parameters $\Delta_1 = \Delta_4 = +1$, $\Delta_2 = \Delta_3 = -1$ to keep track of the commuting / anticommuting character of the fields. The wave equation $(\nabla^2 - m_i^2) \tilde{f}_i = 0$ can be expanded in Rindler modes, $\tilde{f}_i = e^{-i E T} u_{i,E}$, with $u_{i,E}$ satisfying
\begin{equation}
(-\partial_z^2 + V_i(z)) u_{i,E} = E^2 u_{i,E} \, , \qquad V_i(z) = m_i^2 \rho^2 = m_i^2 \rho_0^2 e^{2 z} \, ,
\end{equation}
in terms of a new coordinate $z = \log(\rho/\rho_0)$ and some arbitrary scale $\rho_0$. This potential produces a continuous spectrum of states for $E>0$. In a WKB approximation, the number of states up to energy $E$ is
\begin{equation}
g_i(E) = \int_0^{E/m_i} \frac{\rd \rho}{\pi \rho} \sqrt{E^2 - m_i^2 \rho^2} \, .
\end{equation}
This is of course infinite due to the $\rho \to 0$ UV divergence, we could manually add a regulator but we will see that for the relevant thermodynamic quantities the combination of the original field with the Pauli-Villars regulators already produces a finite result.

Indeed, from the density of states $\rd g_i / \rd E$ we can write the free energy as
\begin{align}
\nonumber \beta F & = \sum_i \Delta_i \int_0^{\infty} \rd E \frac{\rd g_i}{\rd E} \log (1 - e^{-\beta E}) \\
& = - \frac{\beta}{\pi} \int_0^{\infty} \frac{\rd E}{e^{\beta E} - 1} \left( \sum_{i} \Delta_i \int_{0}^{E/m_i} \frac{\rd \rho}{\rho} \sqrt{E^2 - m_i^2 \rho^2} \right) \, ,
\end{align}
where each of the fields contributes to $F$ with the appropriate sign $\Delta_i$. We want to isolate the UV divergence as $\Lambda \to \infty$. The sum of integrals can now be done by setting the lower limit to $\delta$ and taking $\delta = 0$ after combining all terms, since $\sum_i \Delta_i = 0$ guarantees no divergent terms remain. There is an IR logarithmic divergence in the first integral due to the tachyon being massless, $m_1 = 0$. We also found this IR issue in the computation of the effective action, we could regulate it but it is irrelevant for the cancellation of UV divergences we want to see. Keeping then only the divergent term as $\Lambda \to \infty$,
\begin{equation}
F = - \frac{1}{\pi} \int_0^{\infty} \frac{\rd E \, E/2}{e^{\beta E} - 1} \log \left(\frac{m_2^2 m_3^2}{m_4^2} \right) + \ldots = - \frac{\pi}{12 \beta^2} \log (\Lambda^2) + \dots \, .
\end{equation}

The scalar fields in the Rindler wedge are in a thermal state with temperature $\beta = 2\pi$, so their von Neumann entropy agrees with the thermal entropy. We can extract its UV divergent piece from $S = \beta^2 \partial_\beta F$, 
\begin{equation}
S_{\rm mat} = \frac{\pi}{6 \beta} \log (\Lambda^2) + \ldots =  \frac{1}{12} \log (\Lambda^2) + \ldots\, .
\end{equation}
This is exactly the UV divergent contribution to the generalized entropy needed to renormalize the topological coupling,
\begin{equation}
S_{\rm gen} = 4 \pi b_B + 2 \pi a_1 e^{-2\Phi(\phi_R)} + \frac{1}{12} \log (\Lambda^2) + \ldots = 4 \pi b + 2 \pi a_1 e^{-2\Phi(\phi_R)} + \dots \, ,
\end{equation}
thus producing a UV finite result.

\section{Details on the bosonization and nonlocal transformation}
\label{sec:details-bosonization-nonlocal-transformation}
There are some important subtleties regarding the combination of bosonization, nonlocal transformation and WKB limit used in \secref{sec:entanglement-MM}.
The first concerns the bosonization, since different authors use slightly different bosonization procedures.
In this paper, we employ the bosonization conventions of \cite{Klebanov:1991qa} while the conventions for the nonlocal transformation are taken from \cite{Natsuume:1994sp}.
This appendix briefly summarizes the different conventions and explains why those of \cite{Klebanov:1991qa} are more convenient for computing the entanglement entropy by showing that correlation functions are obtained by Wick contractions in the WKB limit.

In the conventions of \cite{Klebanov:1991qa}, standard bosonization rules for a non-relativistic setting are employed.
First introduce left- and right-moving fermions $\psi_\pm$,\footnote{In principle, $\psi_+$ and $\psi_-$ are related by a boundary condition so that the the number of degrees of freedom is unchanged by going over from $\psi$ to $\psi_\pm$. However, in the double-scaling limit the boundary condition is irrelevant \cite{Klebanov:1991qa}.}
\begin{equation}
    \psi(x,t) = \frac{e^{i\mu t}}{\sqrt{2v(x)}}\left[e^{-i \int v(x) \rd x + i\pi/4}\psi_+(x,t) + e^{i\int v(x) \rd x-i\pi/4}\psi_-(x,t)\right]
\end{equation}
where $v(x)$ is defined in \eqref{eq:definition-time-of-flight-variable}. Define now the bosonic field $\bar{S}$ and its canonically conjugate momentum $\bar\Pi$ by
\begin{equation}
    \psi_\pm=\frac{1}{\sqrt{2\pi}} : \exp\left[i\sqrt{\pi}\int \rd\tau (\bar\Pi \mp \partial_\tau \bar S)\right] : \quad \Leftrightarrow \quad \partial_\tau\bar S = \sqrt\pi : \psi^\dagger \psi :\,.
    \label{eq:standard-bosonization}
\end{equation}
This gives the Hamiltonian \eqref{eq:bosonic-Hamiltonian} with standard kinetic term in terms of $\tau$.
Another way to bosonize is to follow \cite{Polchinski:1991uq,Natsuume:1994sp}, which gives a Hamiltonian with standard kinetic term in terms of another variable $q$ defined by $x = -e^{-q}$ (adding the subscript $P$ to distinguish the bosonic field used in the main text from the new field introduced here),
\begin{equation}
  H = \frac{1}{2}\int \rd q \left[\bar\Pi_P^2 - (\partial_q \bar S_P)^2 + \sqrt\pi e^{2q}\left(\bar\Pi_P^2(\partial_q\bar S_P) + \frac{1}{3}(\partial_q \bar S_P)^3\right)\right].
  \label{eq:Hamiltonian-Polchinski}
\end{equation}
These two bosonization procedures are related by a field redefinition $\partial_q \bar S = \partial_q \bar S_P + \frac{1-\sqrt{1-2\mu e^{2q}}}{\sqrt\pi e^{2q}}$.

Both bosonization procedures are valid, neither one is more correct than the other.
However, the description of the WKB limit $\mu \to \infty$, keeping $\tau$ fixed, is much simpler in the conventions of \cite{Klebanov:1991qa} than in those of \cite{Natsuume:1994sp}.
For $\bar{S}$ the WKB limit is simply the non-interacting limit obtained by restricting to the quadratic terms in the Hamiltonian.
As will be shown below, two-point correlators of $\bar S$ computed in the limit $\mu \to \infty$ agree perfectly with those obtained in the WKB limit of the fermionic description if one inserts the bosonization relation \eqref{eq:standard-bosonization}
\begin{equation}
  \ev{\bar S(\tau_1) \bar S(\tau_2)} = -\frac{1}{2\pi}\log\left(\frac{\tau_1-\tau_2}{\tau_1+\tau_2}\right).
\end{equation}
Higher-point correlators of $\bar S$ in the limit $\mu \to \infty$ are simply determined by Wick contractions for all values of $\tau$, as will also be verified below.
On the other hand, for $\bar S_P$, the WKB limit is interacting.
Two-point correlators of $\bar S_P$ computed by dropping the cubic interaction terms in \eqref{eq:Hamiltonian-Polchinski} do not agree perfectly with the fermionic WKB limit,\footnote{Although they do of course agree for $q \to -\infty$, setting $q \approx \tau - 1/2 \log(\mu/2)$, just not in the entire range of $\tau$ where the WKB correlators are valid.}
\begin{equation}
  \ev{\bar S_P(q_1) \bar S_P(q_2)} = -\frac{1}{2\pi}\log\left(\frac{q_1-q_2}{q_1+q_2 + \log(\mu/2)}\right) \neq -\frac{1}{2\pi}\log\left(\frac{\tau_1-\tau_2}{\tau_1+\tau_2}\right).
\end{equation}
Correlators of $\bar S_P$ are only determined by Wick contractions in the $q \to -\infty$ limit.
The difference between the true correlators of $\bar S_P$ and the approximation $\ev{\bar S_P(q_1)\bar S_P(q_2)}_W$ determined by Wick contractions leads to corrections to the target-space two-point correlator.
These corrections are not parametrically suppressed compared to the approximation to the target space correlator $\ev{S_P(\phi_1)S_P(\phi_2)}_W$ determined by Wick contractions.
Therefore, the true correlators in target space are not well approximated by those obtained via dropping the interaction in terms in the bosonization procedure from \cite{Natsuume:1994sp} and thus the methods for the computation of entanglement entropies in free theories from \cite{Peschel2003,Casini:2009sr} are not applicable in that case.

In conclusion, the correct non-interacting limit to choose is the WKB limit $\mu \to \infty$ limit, keeping $\tau$ fixed, as in the main text.
This is most naturally implemented in the conventions chosen in \cite{Klebanov:1991qa}.

\subsection{Correlation functions in the WKB limit}
To put what was said above in words on solid ground, let us now present computations of the two and four-point functions of matrix model scalars in the WKB limit.
These computations illustrate two things.
First, they show that the two-point function obtained by bosonizing a fermionic correlator in the WKB limit is captured exactly by a direct computation of the bosonic two-point function in the ground state, provided one uses the conventions of \cite{Klebanov:1991qa}.
Second, they determine through an explicit computation that the four-point function in the WKB limit factorizes into Wick contractions, proving that interactions can be neglected.

For the two-point function we start with the right hand side of the bosonization relation \eqref{eq:standard-bosonization}.
Integrating and using the Schrödinger equation gives
\begin{equation}
  \bar{S}(t,\tau) = \int_{-\infty}^\infty \rd\nu_1 \rd\nu_2 e^{i(\nu_2-\nu_1)t} a^\dagger_{\nu_1}a_{\nu_2} f(\nu_1,\nu_2,\tau),
\end{equation}
where $a_\nu,a^\dagger_\nu$ are the fermionic creation/annihilation operators $\{a_{\nu_1},a^\dagger_{\nu_2}\} = \delta(\nu_1 - \nu_2)$ and
\begin{equation}
  f(\nu_1,\nu_2,\tau) = \frac{1}{2(\nu_1-\nu_2)}(\psi'_{\nu_1}(\tau)\psi_{\nu_2}(\tau) - \psi_{\nu_1}(\tau)\psi_{\nu_2}'(\tau)).
\end{equation}
Here $\psi_\nu(\tau)$ is the wavefunction solving the time-independent Schrödinger equation
\begin{equation}
  \left(-\frac{1}{2}\partial_\tau^2+V(\tau)\right)\psi_\nu(\tau) = - \nu\, \psi_\nu(\tau).
\end{equation}
The two-point function in the WKB limit is given by
\begin{equation}
  \begin{aligned}
    \ev{\bar{S}(t,\tau_1)\bar{S}(t,\tau_2)} &= \pi \int_{\mu}^\infty \rd\nu_1 \int_{-\infty}^{\mu} \rd\nu_2 f(\nu_1,\nu_2,\tau_1)f(\nu_2,\nu_1,\tau_2).\\
                                      &= \frac{1}{2\pi}\int_0^\infty \rd\delta \int_{-\delta}^\delta \rd\nu \frac{1}{\delta^2} \sin(2\delta \tau(\tau_1))\sin(2\delta \tau(\tau_2))\\
                                      &= -\frac{1}{2\pi} \log\left(\frac{\tau_1-\tau_2}{\tau_1+\tau_2}\right),
  \end{aligned}
  \label{eq:2pt-function-WKB}
\end{equation}
where $\nu_{1,2} = \mu+\nu \pm \delta$.
The bosonic computation of the same quantity uses the expansion \eqref{eq:Fourier-expansion-collective-field} of $\bar{S}$ in terms of creation/annihilation operators.
The bosonic ground state is defined by
\begin{equation}
  \bar{\alpha}_\pm(\omega<0)\ket{0} = 0, \quad \bra{0}\bar{\alpha}_\pm(\omega>0) = 0.
  \label{eq:action-on-ground-state-creation-annihilation-operators}
\end{equation}
 while the commutation relations are given by
 \begin{equation}
  [\bar{\alpha}_\pm(\omega_1),\bar{\alpha}_\pm(\omega_2)] = 2\pi\omega_2\delta(\omega_1+\omega_2).
  \label{eq:commutators-creation-annihilation-operators}
\end{equation}
Expanding in creation/annihilation operators, we get four terms in the two point function $\ev{\bar{S}(t,\tau_1)\bar{S}(t,\tau_2)}$.
The terms with two left or two right-moving operators give
\begin{equation}
  \int_{-\infty}^\infty \frac{\rd\omega_1\rd\omega_2}{(2\pi)^2\omega_1\omega_2} e^{i\omega_1(t \mp \tau_1)+i\omega_2(t \mp \tau_2)} \left(\frac{2}{\mu}\right)^{\mp i(\omega_1+\omega_2)/2}\ev{\bar{\alpha}_\pm(\omega_1)\bar{\alpha}_\pm(\omega_2)} = \int_{-\infty}^0 \frac{\rd\omega}{2\pi\omega} e^{\mp i\omega(\tau_1 - \tau_2)}.
\end{equation}
For the mixed terms containing an expectation value of $\bar{\alpha}_+(\omega_1)\bar{\alpha}_-(\omega_2)$, we use the tree-level operator relation between the left- and right-moving operators \cite{Natsuume:1994sp},\footnote{Note that although our bosonization conventions follow \cite{Klebanov:1991qa} instead of \cite{Natsuume:1994sp}, the operator relation between left- and right-moving creation/annihilation operators is the same in both conventions.}
\begin{equation}
  \bar{\alpha}_-(\omega) = - \left(\frac{2}{\mu}\right)^{-i\omega} \bar{\alpha}_+(\omega) + O(\bar{\alpha}_+^2),
\end{equation}
where $O(\bar{\alpha}_+^2)$ denotes the fact that there is an infinite series of higher order terms containing products of $\bar{\alpha}_+$.
We then find
\begin{equation}
  \int_{-\infty}^\infty \frac{\rd\omega_1\rd\omega_2}{(2\pi)^2\omega_1\omega_2} e^{i\omega_1(t \mp \tau_1)+i\omega_2(t \pm \tau_2)} \left(\frac{2}{\mu}\right)^{\mp i(\omega_1-\omega_2)/2} \ev{\bar{\alpha}_\pm(\omega_1)\bar{\alpha}_\mp(\omega_2)} = -\int_{-\infty}^0 \frac{\rd\omega}{2\pi\omega} e^{\mp i\omega\left(\tau_1 + \tau_2\right)}.
\end{equation}
We thus find basically the same integrals over $\omega$ as we had in the fermionic contribution for integrals over $\delta$ in \eqref{eq:2pt-function-WKB} when expanding the sine functions in terms of exponentials.
The bosonic computation therefore gives the same two-point function,
\begin{equation}
  \ev{\bar{S}(t,\tau_1)\bar{S}(t,\tau_2)} = -\frac{1}{2\pi} \log\left(\frac{\tau_1-\tau_2}{\tau_1+\tau_2}\right).
\end{equation}

For the four-point function $\ev{\bar{S}(\tau_1)...\bar{S}(\tau_4)}$, we would like to show that it factorizes in the WKB limit,
\begin{equation}
  \begin{aligned}
    \ev{\bar{S}(\tau_1)...\bar{S}(\tau_4)} = &\ev{\bar{S}(\tau_1)\bar{S}(\tau_2)}\ev{\bar{S}(\tau_3)\bar{S}(\tau_4)}\\
    + &\ev{\bar{S}(\tau_1)\bar{S}(\tau_3)}\ev{\bar{S}(\tau_2)\bar{S}(\tau_4)}\\
    + &\ev{\bar{S}(\tau_1)\bar{S}(\tau_4)}\ev{\bar{S}(\tau_2)\bar{S}(\tau_3)},
  \end{aligned}
\end{equation}
or equivalently, that the connected four-point function vanishes.
It is somewhat simpler to show this for the derivative of this quantity,
\begin{equation}
  \partial_{\tau_1}...\partial_{\tau_4}\ev{\bar{S}(\tau_1)...\bar{S}(\tau_4)}_\text{conn.} = 0.
\end{equation}
The computation proceeds by first inserting the bosonization relation \eqref{eq:standard-bosonization} as well as the decomposition of the fermion field into creation/annihilation operators.
Then by a straightforward but tedious computation (commuting annihilation operators until they act on the ground state), we obtain
\begin{equation}
  \small
  \begin{aligned}
    &\partial_{\tau_1}...\partial_{\tau_4}\ev{\bar{S}(\tau_1)...\bar{S}(\tau_4)}_\text{conn.}\\
    &= \pi^2 \int_\mu^\infty \rd\nu_1 \rd\nu_4 \int_{-\infty}^\mu \rd\nu_2\rd\nu_3\bigl[\psi_{\nu_1}(\tau_1)\psi_{\nu_1}(\tau_3)\psi_{\nu_2}(\tau_1)\psi_{\nu_2}(\tau_2)\psi_{\nu_3}(\tau_2)\psi_{\nu_3}(\tau_4)\psi_{\nu_4}(\tau_3)\psi_{\nu_4}(\tau_4)\\
    & \qquad + \psi_{\nu_1}(\tau_1)\psi_{\nu_1}(\tau_2)\psi_{\nu_2}(\tau_1)\psi_{\nu_2}(\tau_3)\psi_{\nu_3}(\tau_3)\psi_{\nu_3}(\tau_4)\psi_{\nu_4}(\tau_2)\psi_{\nu_4}(\tau_4)\\
    & \qquad + \psi_{\nu_1}(\tau_1)\psi_{\nu_1}(\tau_4)\psi_{\nu_2}(\tau_1)\psi_{\nu_2}(\tau_3)\psi_{\nu_3}(\tau_2)\psi_{\nu_3}(\tau_4)\psi_{\nu_4}(\tau_2)\psi_{\nu_4}(\tau_3)\\
    & \qquad + \psi_{\nu_1}(\tau_1)\psi_{\nu_1}(\tau_3)\psi_{\nu_2}(\tau_1)\psi_{\nu_2}(\tau_4)\psi_{\nu_3}(\tau_2)\psi_{\nu_3}(\tau_3)\psi_{\nu_4}(\tau_2)\psi_{\nu_4}(\tau_4) \bigr]\\
    &+ \pi^2 \int_\mu^\infty \rd\nu_1 \int_{-\infty}^\mu \rd\nu_2\rd\nu_3\rd\nu_4 \psi_{\nu_1}(\tau_1)\psi_{\nu_1}(\tau_4)\psi_{\nu_2}(\tau_1)\psi_{\nu_2}(\tau_2)\psi_{\nu_3}(\tau_2)\psi_{\nu_3}(\tau_3)\psi_{\nu_4}(\tau_3)\psi_{\nu_4}(\tau_4)\\
    &+ \pi^2 \int_\mu^\infty \rd\nu_1\rd\nu_3\rd\nu_4 \int_{-\infty}^\mu \rd\nu_2 \psi_{\nu_1}(\tau_1)\psi_{\nu_1}(\tau_2)\psi_{\nu_2}(\tau_1)\psi_{\nu_2}(\tau_4)\psi_{\nu_3}(\tau_2)\psi_{\nu_3}(\tau_3)\psi_{\nu_4}(\tau_3)\psi_{\nu_4}(\tau_4).\\
  \end{aligned}
\end{equation}
To compute the energy integrals in the WKB limit, we argue as in \cite{Hartnoll:2015fca}.
First, insert the WKB wavefunctions
\begin{equation}
  \psi_{\nu_1}(\tau)\psi_{\nu_2}(\tau) = \frac{\cos(P_{\nu_1}(\tau)-P_{\nu_2}(\tau)) - \sin(P_{\nu_1}(\tau) + P_{\nu_2}(\tau))}{\pi(x(\tau)^2-\nu_1)^{1/4}(x(\tau)^2-\nu_2)^{1/4}},
\end{equation}
where $P_\nu(\tau) = \int_{\sqrt{2\nu}}^{x(\tau)} \rd x'\sqrt{x'^2-2\nu}$.
The wavefunctions decay rapidly for $\nu < \mu$ while at $\nu > \mu \gg 1$ in the WKB limit, the wavefunctions oscillate rapidly and hence mostly average out in the $\nu$ integrals.
The only slowly oscillating contributions come from the $\nu \approx \mu$ region for the $\cos(P_{\nu_1}(\tau)-P_{\nu_2}(\tau))$ terms, which for large $\nu_1,\nu_2$ are approximated by $\cos\bigl(\tau(\nu_1-\nu_2) + O((\nu_1-\nu_2)^2)\bigr)$.
Therefore, in the WKB limit we set the product of two wavefunctions equal to
\begin{equation}
  \psi_{\nu_1}(\tau)\psi_{\nu_2}(\tau) \to \frac{1}{\pi \sqrt{x(\tau)^2-\mu}}\cos(\tau(\nu_1-\nu_2)).
\end{equation}
Then, the integrals over $\nu_1,...,\nu_4$ can be evaluated explicitly, dropping the endpoints at infinity which represent highly oscillatory contributions that average out.
The result of this procedure is that, as claimed, $\partial_{\tau_1}...\partial_{\tau_4}\ev{\bar{S}(\tau_1)...\bar{S}(\tau_4)}_\text{conn.} = 0$.
Interactions in the bosonic theory are negligible in the WKB limit.

\section{Entanglement for non-invariant algebras}
\label{app:EntanglementNonInvariant}

In this appendix, we will discuss an example which highlights the importance of the algebra of operators chosen to define the entanglement entropy in the presence of symmetries. We consider a quantum mechanical system with Hilbert space $\cH$ on which a symmetry group $G$ acts. There is a subspace invariant under the action of $G$ (the singlet sector), $\cH_0 \subset \cH$, and we will look at an invariant state $\ket{\psi} \in \cH_0$. We will define two algebras of operators $\cA$ and $\cB$ related to the choice of a subsystem with the property that $\cA$ maps $\cH_0$ to itself, while $\cB$ maps $\cH_0$ to $\cH$ (i.e., the image includes non-invariant sectors). If we were treating $G$ as a gauge symmetry, only $\cA$ would be a valid gauge-invariant algebra, but here we are thinking of $G$ as a global symmetry. The goal is to obtain the von Neumann entropy associated to the algebras $\cA$ and $\cB$ and compare them.

Rather than discussing this in a fully abstract language, we focus on a model inspired by the setup of identical particles used to introduce target space entanglement in \cite{Mazenc:2019ety}. Let $\cH$ be the Hilbert space of $N$ (in principle distinct) particles moving on a $d$-dimensional space. We have an action of the permutation group $G = S_N$, and the sector invariant under this action is the subspace of fully symmetric wavefunctions $\cH_0 \subset \cH$. This is the space of wavefunctions of $N$ identical bosons.\footnote{All the discussion that follows can be repeated for identical fermions by replacing $\cH_0$ by the subspace of $\cH$ carrying the sign representation of the permutation group.} In the spirit of the target space entanglement proposal \cite{Mazenc:2019ety}, imagine we divide the space into two complementary regions $R$ and $\bar{R}$. We want to associate an algebra of operators to region $R$, and compute its von Neumann entropy on the state $\ket{\psi}$. We will do this in two ways.

If we work on $\cH_0$ (i.e., if we treat the particles as indistinguishable), we can define the algebra $\cA$ by starting from one-particle operators localized in $\cR$
\begin{equation}
\cO_R = \int_R \rd \mathbf{x} \, \rd\mathbf{y} \, O_R(\mathbf{x}, \mathbf{y}) \ket{\mathbf{x}} \bra{\mathbf{y}} \, ,
\end{equation}
and considering symmetrized combinations
\begin{equation}
\cO_R^{\rm (sym)} =\frac{1}{N} \left( \cO_{R} \otimes \mathbf{1} \otimes \dots \otimes \mathbf{1} + \mathbf{1} \otimes \cO_{R} \otimes \dots \otimes \mathbf{1} + \ldots + \mathbf{1} \otimes \mathbf{1} \otimes \dots \otimes \cO_{R} \right) \, .
\end{equation}
The algebra $\cA$ is generated by linear combinations and products of operators $\cO_R^{\rm (sym)}$ together with the identity -- note that operators acting simultaneously on more than one particle arise from products of the operators above. These operators manifestly leave $\cH_0$ invariant, since they act democratically on all the $N$ particles.

Computing the von Neumann entropy of this algebra requires some care because $\cH_0$ has superselection sectors -- subspaces invariant under the action of $\cA$. The general discussion can be found in \cite{Mazenc:2019ety}, but essentially we can write
\begin{equation}
\cH_0 = \bigoplus_{k=0}^N \cH_{0,k} \, ,
\end{equation}
where $\cH_{0,k}$ is the subspace of (invariant) states in which $k$ particles are in region $R$ and $N-k$ particles are in region $\bar{R}$. The algebra $\cA$ does not mix sectors with different values of $k$, and for a particular $k$ it acts only on the particles in region $\cR$. At the level of the algebra $\cA$, we can thus view $\cH_{0,k}$ as a tensor product $\cH^{(R)}_k \otimes \cH^{(\bar{R})}_{N-k}$, with $\cA$ acting only on the first factor corresponding to $k$ identical particles in region $R$ (the second factor corresponds to $N-k$ particles in region $\bar{R}$).  Within each sector $\cH_{0,k}$, we thus define a density matrix by tracing out the $\cH^{(\bar{R})}_{N-k}$ factor
\begin{align}
\nonumber \tilde{\rho}_{0,k}(\mathbf{x}_1, \dots, \mathbf{x}_k; \mathbf{y}_1, \dots \mathbf{y}_k) = \binom{N}{k} \int_{\bar{R}} \rd \mathbf{z}_1 \dots \rd \mathbf{z}_{N-k} & \, \psi (\mathbf{x}_1, \dots, \mathbf{x}_k, \mathbf{z}_1, \dots, \mathbf{z}_{N-k}) \\
& \times \psi^{\star} (\mathbf{y}_1, \dots, \mathbf{y}_k, \mathbf{z}_1, \dots, \mathbf{z}_{N-k}) \, .
\end{align}
The binomial coefficient arises because of the number of ways to pick $k$ particles among $N$. Note that this is an unnormalized density matrix, in the sense that
\begin{equation}
p_{0,k} \equiv {\rm tr}_{\cH_k^{(R)}} \left( \tilde{\rho}_{0,k} \right) \neq 1 \, ,
\end{equation}
where the trace is over the space of $k$ identical particles in $R$. This $p_{0,k}$ is actually the probability to find $k$ particles in region $R$: as a consistency check of all normalizations, one can prove that $\sum_k p_{0,k} = 1$. The von Neumann entropy for the algebra $\cA$ is defined as the sum of the entropies for each sector \cite{Mazenc:2019ety}
\begin{equation}
S(\cA) = - \sum_k {\rm tr}_{\cH_k^{(R)}} \left( \tilde{\rho}_{0,k} \log \tilde{\rho}_{0,k} \right) = - \sum_k p_{0,k} \log p_{0,k} + \sum_k p_{0,k} \, S(\rho_{0,k}) \, ,
\end{equation}
where in the last step we have defined the normalized density matrix $\rho_{0,k} = \tilde{\rho}_{0,k}/p_{0,k}$, and the final $S(\rho_{0,k})$ is the standard von Neumann entropy of this state.

We can adopt an alternative perspective if we view the particles as distinguishable, i.e., if we consider the \emph{same} invariant state $\ket{\psi}$ but as part of the full Hilbert space of $N$ particles $\cH$. In that case, the natural way to define the algebra of operators $\cB$ corresponding to region $R$ is by using linear combinations and products of the identity and operators of the form
\begin{equation}
\mathbf{1} \otimes \cdots \otimes \mathbf{1} \otimes \mathbf{\cO}_R \otimes \mathbf{1} \otimes \cdots \otimes \mathbf{1} \, ,
\end{equation}
where the $\cO_R$ can be in any of the $N$ positions. Now the space $\cH$ splits into many more superselection sectors, labeled by which particles are in region $R$ and which in region $\bar{R}$ (where we can now differentiate particles whose labels are in different positions of the wavefunction). For notational convenience, define $\sigma$ to be one of the $2^N$ strings of $N$ characters $R$ and $\bar{R}$, with the character in position $k$ denoting in which region the $k$-th particle is. Then the Hilbert space splits as
\begin{equation}
\cH = \bigoplus_{\sigma} \cH_{\sigma} \, ,
\end{equation}
and each term can be seen as a tensor product
\begin{equation}
\cH_{\sigma} \cong \cH_{\sigma}^{(R)} \otimes \cH_{\sigma}^{(\bar{R})} \, .
\end{equation}
The first factor collects the $k$ particles in region $R$ (where $k$ is equal to the number of characters $R$ in $\sigma$), while the second collects the $N-k$ particles in $\bar{R}$. The algebra $\cB$ acts non-trivially only on the first factor.

Much like for the algebra $\cA$, we can define an unnormalized density matrix for $\cB$ within each sector
\begin{align}
\nonumber \tilde{\rho}_{\sigma}(\mathbf{x}_1, \dots, \mathbf{x}_k; \mathbf{y}_1, \dots \mathbf{y}_k) = \int_{\bar{R}} \rd \mathbf{z}_1 \dots \rd \mathbf{z}_{N-k} & \, \psi (\sigma[\mathbf{x}_1, \dots, \mathbf{x}_k; \mathbf{z}_1, \dots, \mathbf{z}_{N-k}]) \\
& \times \psi^{\star} (\sigma[\mathbf{y}_1, \dots, \mathbf{y}_k; \mathbf{z}_1, \dots, \mathbf{z}_{N-k}]) \, ,
\end{align}
where $\sigma[\mathbf{x}_1, \dots, \mathbf{x}_k; \mathbf{z}_1, \dots, \mathbf{z}_{N-k}]$ puts the $\mathbf{x}_i$ labels in the positions containing $R$ in $\sigma$ and the $\mathbf{z}_i$ labels in those containing $\bar{R}$. The entropy of the algebra $\cB$ is then
\begin{equation}
S(\cB) = - \sum_\sigma {\rm tr}_{\cH_\sigma^{(R)}} \left( \tilde{\rho}_{\sigma} \log \tilde{\rho}_{\sigma} \right) = - \sum_\sigma p_{\sigma} \log p_{\sigma} + \sum_\sigma p_{\sigma} \, S(\rho_{\sigma}) \, ,
\end{equation}
where
\begin{equation}
p_\sigma = {\rm tr}\left( \tilde{\rho}_{\sigma} \right) \, , \qquad \rho_{\sigma} = \frac{\tilde{\rho}_{\sigma}}{p_{\sigma}} \, ,
\end{equation}
are respectively the probability for the particles to be distributed as in $\sigma$ and the corresponding normalized state in $\cH_{\sigma}^{(R)}$.

For a general state, $S(\cB)$ and $S(\cA)$ are in principle unrelated. However, since we are looking at invariant states, we can find a relation between them. The total symmetry of the wavefunction implies that\footnote{$\tilde{\rho}_{0,k}$ and $\tilde{\rho}_{\sigma}$ live a priori in different spaces: that of $k$ identical particles in $R$ and that of $k$ distinguishable particles in $R$. The former is a subspace of the latter, and since the state is completely invariant, $\tilde{\rho}_{\sigma}$ can actually be seen as living in the space of $k$ identical particles in $R$.}
\begin{equation}
\tilde{\rho}_{0,k} = \binom{N}{k} \, \tilde{\rho}_{\sigma} \, , \qquad p_{0,k} = \binom{N}{k} \, p_{\sigma} \, , \qquad \rho_{0,k} = \rho_{\sigma} \, .
\end{equation}
Plugging this in the definition of both entropies, we get a relation
\begin{equation}
\label{EntanglementNonInvariant:RelationEntropies}
S(\cB) = S(\cA) + \sum_{k=0}^N p_{0,k} \log \binom{N}{k} \, ,
\end{equation}
which is the result quoted in the main text. We see then how the different algebras give rise to different results for the entropy, but at the same time there is a simple relation between them because of the symmetry of the state considered.

\bibliographystyle{JHEP}
\bibliography{bibliography.bib}

\end{document}